\documentstyle[floats,twocolumn,aps,prl,epsf,rotate]{revtex}

\def\gtrsim{\compoundrel>\over\sim}
\def\lesssim{\compoundrel<\over\sim}
\def\compoundrel#1\over#2{\mathpalette\compoundreL{{#1}\over{#2}}}
\def\compoundreL#1#2{\compoundREL#1#2}
\def\compoundREL#1#2\over#3{\mathrel
     {\vcenter{\hbox{$\buildrel{#1#2}\over{#1#3}$}}}}

\draft
\tighten

\begin{document}

\title{Reactions at surfaces studied by {\it ab initio}
dynamics calculations}

\author{Axel Gro{\ss}\cite{newaddr}\\[.5cm]
Fritz-Haber-Institut der Max-Planck-Gesellschaft, Faradayweg 4-6, 
D-14195 Berlin-Dahlem, Germany
}

\maketitle

\begin{abstract}

Due to the development of efficient algorithms and the improvement of computer
power it is now possible to map out potential energy surfaces (PES)
of reactions at surfaces in great detail. This achievement has been 
accompanied by an increased effort in the dynamical simulation 
of processes on surfaces. The paradigm for simple reactions
at surfaces -- the dissociation of hydrogen on metal surfaces --
can now be treated fully quantum dynamically in the molecular degrees
of freedom from first principles, i.e., without invoking any adjustable 
parameters. This relatively new field of {\it ab initio} dynamics
simulations of reactions at surfaces will be reviewed.  Mainly the
dissociation of hydrogen on clean and adsorbate covered metal surfaces
and on semiconductor surfaces will be discussed.
In addition, the {\it ab initio} molecular dynamics treatment of 
reactions of hydrogen {\em atoms} with hydrogen-passivated semiconductor
surfaces and recent achievements in the {\it ab initio} description of 
laser-induced desorption and further developments will be addressed.

\end{abstract}

\tableofcontents

\section{Introduction}

Understanding reactions on surfaces plays an important role in a wide range
of technologically relevant applications. Among those are
the {\em heterogenous catalysis} -- the majority of reactions in the
chemical industry employ catalysts;
{\em crystal growth}, which determines, e.g., the quality of
semiconductor devices; {\em corrosion} and {\em lubrication}, which
influences the durability of mechanical systems;
or {\em hydrogen storage} in metals, just to mention
a few. The reactions involved in these processes are often too complicated
to be studied in detail as a whole. Therefore in surface science one
tries to understand reaction mechanisms by breaking them up 
into simpler steps which are then studied under well-defined
conditions \cite{Zan88,Des96}.

Studies of chemical reaction dynamics are well established in the gas phase. 
Reactions on surfaces differ from gas-phase reactions in two fundamental 
aspects: first, the presence of the surface changes the symmetry significantly,
and second, the substrate on which the reaction occurs represents
in principle a system with an infinite number of degrees of freedom which acts 
as a heat bath that leads to dissipation and thermal fluctuations. 
As for the symmetry, while for example for the description of the dissociation
of a diatomic molecule in the gas phase only the interatomic distance has to be
taken into account (all other degrees of freedom can be separated),
in front of a surface all six degrees of freedom of the diatomic molecule  
have to be considered explicitly since the surface breaks the symmetry
with respect to the gas phase. On the other hand, a crystalline surface
introduces new symmetries, for example the periodicity of the surface.

Dissociative adsorption processes are of particular importance for reactions 
on surfaces since they constitute the first step in heterogenous catalysis; 
furthermore they are often the rate-limiting process, for example in ammonia 
synthesis or CO oxidation. The term ``heterogenous'' refers to the fact that 
the reactants (usually a gas) and the catalyst (usually a solid substrate) are
not in the same phase state.
For the atomic or molecular adsorption process, energy dissipation to the 
substrate is necessary for the particles to be trapped into an adsorption well
on the surface, otherwise they would
be scattered back into the gas phase. In the case of dissociative
adsorption, however, there is another channel for energy transfer, which
is the conversion of the kinetic and internal energy of the molecule into 
translational energy of the atomic fragments on the surface relative 
to each other. In addition,
for light molecules like hydrogen  there is usually only a small energy
transfer to substrate phonons due to the large mass mismatch between the
molecule and the substrate atoms. If furthermore no substantial surface
rearrangement upon adsorption occurs -- which is often fulfilled for densely 
packed metal surfaces -- then the dissociative adsorption dynamics 
can be described using low-dimensional potential energy surfaces (PES).

Still the interaction of a diatomic molecule with a well-defined 
fixed substrate involves the six degrees of freedom 
of the molecule. Up to recently the interaction dynamics in particular
of hydrogen had been described in low-dimensional quantum studies
\cite{Jac87,Hal90,Dar92a,Kue91,Sch92,Dar92b,%
Bre94a,Han90,Gro94e,Dar94a,Mow93,Dar94JCP,Bru94,Dai95,%
Kar87,Nie90,Hal88,Dar90,Dar92c,Dar94,%
Gro93CP,Gro93SS,Gro94SS,Gro95JCP,Din97};
higher-dimensional studies could only be performed by classical
trajectory calculations  \cite{Kara90,Yang90,Eng93CPL}
or mixed quantum-classical methods \cite{Eng92,Eng93,Gru93,Kum94}
on model potentials.
This was due to the lack of reliable potential energy surfaces and the 
large computational effort for high-dimensional quantum
dynamics studies (for reviews on these dynamical studies see, e.g., 
Refs.~\cite{DeP91,Hol94,Tul94,Dar95rep}). 
This situation has changed significantly within the
last five years caused by the development of efficient algorithms
and the increase in computer power. For the paradigm of simple reactions
on surfaces -- the dissociation of hydrogen on metal surfaces --
detailed potential energy surfaces obtained by density functional
theory calculations are now available 
\cite{Ham94,Whi94,Wil95,Wil96PRB,Wil96S,Wei97,Ham95PRL,%
          Whi96PRB,Wie96,Eich96,Dong96}.
This has increased the motivation to perform dynamical studies
in which all hydrogen degrees of freedom are treated explicitly,
and indeed the first six-dimensional quantum dynamical studies
of hydrogen dissociation on metal surfaces have now been performed
\cite{Gro95PRL,Gro98PRB,Kro97PRL,Kro97JCP}.

In this review I will give a overview of this rather new field  of 
{\it ab initio} dynamics calculations of reactions on surfaces.
Usually these calculations require three independent steps:\\[-.8cm]
\begin{enumerate}
\item Determination of the {\it ab initio} PES by first-principles 
      total-energy calculations,
\item a fit of the total energies to an analytical or numerical 
      continuous representation 
      which serves as an interpolation between the actually calculated points,
\item a dynamical calculation on this representation of the {\it ab initio} PES
      that includes all relevant degrees of freedom.\\[-.8cm]
\end{enumerate}
Since this type of calculations is indeed derived from first principles 
with no adjustable parameters, I will refer to them as  
``{\it ab initio} dynamics calculations''. However, this 
term should not conceal the still approximative nature of these dynamics
simulations. In all three steps approximations are involved
which will be discussed in this review. For example, 
exchange and correlation effects of the electrons can not be treated exactly
in total-energy calculations using density functional theory.
The description of the surface by either a supercell or a cluster method
is an approximation. Furthermore, the fitting of the {\it ab initio}
energies to a continuous representation is a highly non-trivial task.
If just classical trajectories are determined, quantum effects in the motion 
of the nuclei are neglected which can be important, in particular 
for the dynamics of hydrogen.
And also the question, what are the relevant degrees of freedom in the
dissociation process, cannot be answered unambiguously and depends, for 
example, on the properties that one is interested in. Hence in this review 
also investigations, in which not all molecular degrees of freedom
are treated explicitly, but which are still derived from first principles,
are included.

For classical {\it ab initio} dynamics simulations, which will in the 
following be referred to by the term ``{\it ab initio} molecular dynamics'',
the three steps mentioned above can actually be combined since most 
first-principles total-energy schemes also determine the gradients of
the potential via the Hellmann-Feynman theorem \cite{Hell37,Feyn39}.
Thus the classical equations of motion can be directly solved.
For each step of the numerical integration of the equations of motion 
the forces are determined by a new total-energy calculation. Since this
is rather time-consuming, only a small number ($\lesssim 100$) 
of trajectories can be calculated in this way. Therefore this  method
does not allow the determination of reaction probabilities which usually
require at least thousands of trajectories due to the statistical
nature of the reactive events \cite{Gro98PRB}. 

On the other hand, {\it ab initio} molecular dynamics simulations with the 
determination of the forces ``on the fly'' do not require any
fitting. This makes them very flexible. On semiconductor surfaces, the
substrate rearrangement upon adsorption cannot be neglected; indeed
it plays a crucial role for the adsorption and desorption 
mechanism \cite{Bre94a,Kol95}.
An analytical fit of an {\it ab initio} PES including several substrate 
degrees of freedom has not been performed yet because of its complexity. 
Therefore, for the investigation of the dissociative adsorption and 
associative desorption process on semiconductor surfaces the 
``traditional'' {\it ab initio} molecular dynamics method 
with the determination of the forces ``on the fly''
has still been used \cite{DeVita93,Gro97H2Si,Silva97,Rad97Rev}. 
Its application is reasonable
if the information obtained from a small number of trajectories is
sufficient to gain insight into a particular process.

High-dimensional {\it ab initio} dynamics calculations will not make
low-dimensional simulations on model potentials obsolete. These
model calculations have provided us with the framework to interprete
the dynamics. High-dimensional dynamics is often too complicated to
be followed in detail. But this review will show that high-dimensional 
{\it ab initio} dynamics calculations have advanced our understanding of the
dissociation dynamics tremendously and sometimes even caused the modification
of established concepts. First of all, the {\it ab initio}
calculation of potential energy surfaces has confronted the gas-surface
dynamics community with some features of these PESs which had not been
expected and which challenged new interpretations and calculations.
And secondly, some phenomema in the reaction dynamics on surfaces 
only occur in simulations if a sufficiently large number of degrees 
of freedom is included. Thus high-dimensional simulations not only lead
to progress in the quantitative, but also in the qualitative understanding
of processes on surfaces. In particular for the hydrogen dissociation
on transition metals surfaces {\it ab initio} dynamics calculations have 
proven their power. They demonstrated, e.g., that the initial decrease of the
sticking probability with kinetic energy often found experimentally 
in these systems is not due to a precursor mechanism, as was the common
believe, but is caused by an hitherto underestimated dynamical mechanism, 
namely the steering effect \cite{Gro95PRL,Kay95}.

This review is devoted to a large part to the dissociation of hydrogen
on surfaces. From the theoretical point of view, hydrogen is the
simplest molecule. This does not neccessarily keep the computational effort 
small, but it makes the theoretical description controllable. From the
experimental point of view, hydrogen molecules can be used without severe
problems in molecular beam apparatuses, and they can be detected 
state-specifically by laser techniques. Therefore hydrogen is the ideal
candidate for a close collaboration between theory and experiment in
order to explore the dynamics of the dissociation process on surfaces.
The general concepts found in these studies will also be applicable
to heavier and more complicated molecules. Due to its light
mass, the study of the hydrogen dissociation dynamics also allows
to address some fundamental concepts in physics like the importance
of quantum effects. Still the hydrogen interaction with surface is 
technologically relevant, for example in the passivation or growth of 
semiconductor devices. All these facts make the study of the hydrogen 
dissociation dynamics on surfaces an exciting research subject. 

The electronically adiabatic {\it ab initio} dynamics of adsorption and
desorption has recently been reviewed from a more quantum chemical 
point of view \cite{Rad97Rev}. In the meantime, 
an {\it ab initio} molecular dynamics study of the
reaction of hydrogen {\em atoms} with a hydrogen-passivated semiconductor
surface has been performed \cite{Kra98} in which pick up reactions
of the Eley-Rideal type were studied. 
Furthermore, detailed {\it ab initio} PES studies of reactions on surfaces
involving heavier molecules like O$_2$ \cite{Eich97PRL,Gra96PRL,Bird97} or 
CO$_2$ \cite{Eich98Diss,Sta97PRL}
have been performed, and probably  soon dynamics calculations using
these {\it ab initio} potentials will be carried out. Even the laser-induced
desorption which involves electronically excited states has currently been
addressed from first principles, namely the laser-induced desorption of NO 
form NiO(100) \cite{Klue97,Klue98PRL}, which will also be briefly reviewed
at the end of this article. This short collection shows that 
{\it ab initio} dynamics studies of reactions on surface is a very 
active and growing field. 

After this introduction, in the next section the theoretical concepts 
necessary in order to perform {\it ab initio} dynamics calculations are
briefly introduced. Then the hydrogen dissociation on metal surfaces
is addressed, first on simple and noble metals where the dissociation 
is usually hindered by a substantial barrier, and then on transition
metals which often have non-activated pathways to dissociative
adsorption. In this context also the poisoning of the dissociation
by an adsorbate is discussed. The interaction of hydrogen with
semiconductor surfaces is treated in the following section.
Then the first {\it ab initio} dynamics calculation involving
electronically excited states is discussed.
The review ends with an outlook and some concluding remarks.

\section{Theoretical Background}
In this section the theoretical methods to describe the molecule-surface
interaction are briefly described. It is not intended to give a complete
overview. I will rather introduce the main concepts and shortly comment
on them.

\subsection{Born-Oppenheimer Approximation}
The Schr{\"o}dinger equation describing the interaction of molecules with
surfaces has the general form
\begin{equation}\label{GenSch}
{\cal H} \ | \Psi(\{{\bf R}_m, {\bf r}_n\})  \rangle \ =
\ {\cal E} \ | \Psi(\{{\bf R}_m, {\bf r}_n\})  \rangle.
\end{equation} 
In this equation ${\bf R}_m$ are the
ionic coordinates and ${\bf r}_n$ the electronic coordinates.
It is well-known that a complete analytical solution of the Schr{\"o}dinger
equation taking into account both ionic and electronic degrees of
freedom is not possible except for simple cases. One common approach
is to assume that -- due to the large mass difference between electrons
and the nuclei -- the electrons follow the motion of the nuclei
adiabatically. This is the famous Born-Oppenheimer approximation
\cite{Born27}. And since it is only an approximation, its validity
has to be checked carefully. 

In gas-surface scattering electronically non-adiabatic processes are 
indeed occuring. They are directly observable as 
chemiluminescence and exo-electron emission (see, e.g., the recent
review of T. Greber \cite{Gre97}). However, the proper treatment of
electronically non-adiababic effects in gas-surface dynamics is
rather complex. One problem that goes with the complexity of the
non-adiabaticity is that it is not easy to judge whether the dynamics
in any particular system is indeed electronically adiabatic or not.
The usual argument for the validity of the Born-Oppenheimer approximation
is the smallness of the atomic velocities as compared to the electronic
velocities. If in addition the adiabatic energy levels are well-separated,
electronic transitions are negligible (see, e.g., Ref.~\cite{Born54}).
For metals the situation is more complicated due to the quasi-continuum
of electronic states. Since the effective potentials and the coupling
between the electronic states can still not be computed rigorously,
one is left with more or less hand-waving arguments. Among them are the
following: 
(i)  At metal surfaces electronic excitations are very short-lived. Hence
electronic excitations are effectively quenched.
(ii) Molecular electronic levels become rather broad upon the interaction
with surfaces. Broad levels correspond to short lifetimes of excited states
and again, this leads to an effective quenching of electronic excitations.

On semiconductor and insulator surfaces the situation is different
due to the band gap which makes the treatment of electronically excited
states more tractable. 
At the end of this review I will present the first description of 
electronically non-adiabatic processes in molecular desorption from 
oxide surfaces on an {\it ab initio} basis. 
For metal surfaces, where no band gap exists,
a more practical approach has still to be applied which
is just to perform electronically adiabatic
dynamical studies of reactions on surfaces as well and detailed as
possible and to compare the results of these high-quality
calculations with experiment. As long as the 
consequences of these simulations are in agreement with experiment,
there is apparently no need for the involvement of electronic excitations
in the described processes.
For the time being, we just follow this practical approach and
assume that the Born-Oppenheimer approximation is justified.
We can then write down the electronic Hamiltonian in which the coordinates
of the nuclei just enter as parameters. This Hamiltonian has the form
\begin{equation}
H_{el} \ (\{ {\bf R}_m \}) \ | \psi (\{{\bf r}_n\}) \rangle \ = 
\ E \ (\{ {\bf R}_m \}) \ | \psi (\{{\bf r}_n\}) \rangle .
\end{equation}
The many-electron ground state energy  $E_0 (\{ {\bf R}_m \} )$ then
defines the potential for the motion of the nuclei. Once it is obtained,
it can be plugged into the Schr{\"o}dinger equation for the
nuclei,
\begin{equation}\label{Schr}
\left( \sum_i \frac{-\hbar^2}{2M_i} \nabla^2_{{\bf R}_i} \  + 
      \ E_0 (\{ {\bf R}_m \} ) \right)  \ \Phi(\{{\bf R}_m\}) =
      \ {\cal E} \ \Phi(\{{\bf R}_m\}),
\end{equation}
where ${\cal E}$ is now the energy relevant for the dynamics of the nuclei,
or, alternatively, it can be used to solve the classical equations of motion,
\begin{equation}
M_i \ \frac{\partial^2}{\partial t^2} {\bf R}_i \ = \
- \frac{\partial}{\partial {\bf R}_i} E_0 (\{ {\bf R}_m \} ) \quad .
\end{equation}
For extended systems the most efficient approach to determine the 
many-electron ground state energy $E_0 (\{ {\bf R}_m \} )$ from 
first principles is density functional theory
\cite{Hoh64,Koh65} in combination with the supercell concept. 
In this approach the surface is modelled by a periodic array of slabs
separated by vacuum regions, and the wave functions are expanded in a
suitable basis which is often based on plane waves or augmented plane waves. 
Almost all of the {\it ab initio} potentials discussed later in this 
review are determined in this way, 
but real-space \cite{Briggs96,Gro91PM} and Green function 
methods \cite{Schef91,Feib92PRB,Tri96} have been proposed as well.
And also quantum chemical methods are used in which the infinite substrate 
is modelled by a finite cluster (see, e.g., the recent review of Whitten and 
Yang \cite{Whi96}). {\it Ab initio} total energy calculations are briefly
discussed in the next section.

\subsection{{\it Ab initio} total energy calculations}

In this review I will focus on the {\em reaction dynamics on {\em ab initio}
potential energy surfaces}, not on the determination of the PESs. Still 
I will introduce some basics about Density Functional Theory (DFT). For 
further details I refer to more extensive treatments like, e.g., 
Refs.~\cite{Drei90,Pay92}. DFT is based upon the Hohen\-berg-Kohn theorem
\cite{Hoh64} which states that the ground-state total energy of a system of
interacting electrons $E_{\rm tot}$ can be obtained by minimizing an energy 
functional $E[n]$,
\begin{equation}
E_{\rm tot} \ = \ \min E[n] \ = \ \min (T[n] \ + \ U[n] \ + E^{\rm xc}[n]).
\end{equation}
$T[n]$ and $U[n]$ are the functionals of the non-interacting many electron
kinetic and electrostatic energy, respectively. All quantum mechanical
many-body effects are contained in the so far unknown exchange-correlation 
functional $E^{\rm xc}[n]$. The electron density $n({\bf r})$ which minimizes
the total energy can be found by solving self-consistently the 
Kohn-Sham equations \cite{Koh65}
\begin{equation}
H \psi_i ({\bf r}) \ = \ \left[ \frac{-\hbar^2}{2m} \nabla^2 \   
                      + \ V^{\rm es} ({\bf r}) \ + 
                       \ V^{\rm xc} ({\bf r}) \right] \psi_i ({\bf r}) \ 
                     = \ \varepsilon_i \psi_i ({\bf r}),
\end{equation}
where $T$ is the single-particle kinetic energy operator 
and $V^{\rm es} ({\bf r})$ the electrostatic potential. The 
exchange-correlation potential $V^{\rm xc} ({\bf r})$ is the functional
derivative of the exchange-correlation functional $E^{\rm xc} [n]$
\begin{equation}
V^{\rm xc} ({\bf r}) \ = \ \frac{\delta E^{\rm xc} [n]}{\delta n}.
\end{equation}
The exchange-correlation functional $E^{\rm xc} [n]$ can be written as
\begin{equation}
E^{\rm xc} [n] \ = \ \int d^3{\bf r} \ n({\bf r}) 
\ \epsilon^{\rm xc}[n]({\bf r}),
\end{equation} 
where $\epsilon^{\rm xc}[n]$ is the exchange-correlation energy per particle. 
This exchange-correlation energy is not known exactly except for some 
special cases like constant electron density. In a wide range of bulk and
surface problems the so-called local density approximation (LDA), in
which $\epsilon^{\rm xc}[n]$ is replaced by the exchange-correlation energy
for the homogeneous electron gas, has been surprisingly successful
\cite{Pay92}. For dissociation barriers on surfaces, however, LDA is
seriously in error. In the generalized gradient approximation (GGA)
also the gradient of the density is included in the exchange-correlation
functional. Different forms for the GGA functional have been proposed
(see, e.g., Refs. \cite{Becke88,Lee88,Per92,Per96}). The use of GGA 
functionals leads to a significant improvement in the accuracy of
calculated barrier heights \cite{Ham94,Per92}. Still the validity of the
GGA is strongly debated, in particular in the H$_2$/Si(100) system
\cite{Nac96}. The problem in the development of more accurate 
exchange-correlation functionals is that they still represent in principle 
an uncontrolled approximation, i.e., there is no systematic way of improving
the functionals since there is no expansion in some controllable
parameter (it should be noted here that even if an expansion in some 
controllable parameter exists, this expansion can still be problematic). 
Basically the success justifies the choice of some
particular functional. Hence dynamical simulations on {\it ab initio}
potentials also serve the purpose to check the accuracy of the chosen
exchange-correlation functional.

The supercell approach allows to transfer DFT
algorithms, that have been very successfully applied to the determination
of bulk properties, to the description of surface problems.
Especially the seminal paper by Car and Parrinello \cite{Car85}
has drawn a lot of attention. Due to the variational principle of
density functional theory the determination of the electronic ground
state can be regarded as a global optimization scheme thus avoiding
the explicit diagonalization of huge matrices. Such an approach was first
proposed by Bendt and Zunger \cite{Ben83}. Car and Parrinello demonstrated
how powerful this method can be by combining DFT and molecular dynamics.

In the quantum chemical approach the infinite substrate is represented by a
finite cluster. DFT methods have also been used to determine the 
total energies of the cluster \cite{Dun91,Doren96}, but traditionally 
quantum chemical methods based on the wave function dominate, 
i.e. Hartree-Fock methods without
or with the inclusion of the configuration interaction (CI). While supercell
calculations often describe relatively high-coverage situations due to the 
repeated surface unit cell, cluster methods represent low-coverage situations.
However, they suffer from the fact that the infinite substrate is often not
appropriately modelled by a small cluster and that the total energies are
often not converged as a function of the cluster size \cite{Whi96}. 
For the treatment of electronic excitations, DFT methods could in principle
also be used because due to the Hohenberg-Kohn theorem the electronic ground
state density uniquely defines the external potential \cite{Hoh64} and 
thereby also all electronically excited states. However, in practice
DFT methods usually only determine the total energy for the electronic 
ground state of a certain configuration (there are exceptions, see
for example Ref.~\cite{Pan95}). To describe electronically excited states 
wave-function based methods are still unavoidable which necessitates
the use of a finite cluster that can, however, be embedded for example 
in a field of point charges \cite{Klue97,Klue98PRL}.

\subsection{Parametrization of the {\it ab initio} potential}

One serious problem arises for the use of {\it ab initio} potential energies 
in particular in quantum dynamics simulations. 
To solve the Schr{\"o}dinger equation, one needs in general a continuous
description of the potential since the wave functions are delocalized.
The {\it ab initio} calculations, however, just provide total energies
for discrete configurations of the nuclei. In order to obtain a
continuous description, the {\it ab initio} energies have to be fitted
to an analytical or numerical continuous representation of the
potential energy surface. This is a highly non-trivial task. On the one hand
the representation should be flexible enough to accurately reproduce
the {\it ab initio} input data, on the other hand it should have
a limited number of parameters so that it is still controllable.
Ideally a good parametrisation should not only accurately interpolate between 
the actually calculated points, but it should also give a reliable 
extrapolation to regions of the potential energy surface that have 
actually not been determined by the {\it ab initio} calculations.

The explicit form of the chosen analytical or numerical representation
of the {\it ab initio} potential varies from application to application.
Often the choice is dictated by the dynamics algorithm in which the
representation is used. The applications have almost entirely been devoted
to the interaction of a diatomic molecule with the surface. The 
angular orientation of the molecule has usually been expanded in
spherical harmonics and the center-of-mass coordinates parallel to
the surface in a Fourier series \cite{Wie96,Gro95PRL,Kay95,Gro94PRL}.
For the PES in the plane of the molecular distance from the surface and 
the interatomic separation a representation in reaction path coordinate
has been employed \cite{Gro95PRL,Kay95,Gro94PRL,Kin96}, but also two-body
potentials have been used \cite{Wie96}. Before detailed {\it ab initio} 
potentials became available, the LEPS form was often used to construct
a global PES \cite{DeP91}. This parametrization contains only a small
number of adjustable parameters which made it so attractive for model
calculations, but which makes it at the same time relatively unflexible.
A modified LEPS potential has still been successfully used to fit 
an {\it ab initio} PES of the interaction of atomic hydrogen with the
the hydrogenated Si(100) surface \cite{Kra98}.

As an alternative approach to fit {\it ab initio} energies, a genetic 
programming scheme has recently been proposed, in which both the best 
functional form and the best set of parameters are searched for \cite{Mak98}.
This method has so far only been used for three-dimensional potentials
so that its capability still has to be proven for higher-dimensional problems. 

{\it Ab initio} total energies are often mainly determined at high-symmetry
points of the surface in order to reduce the computational cost. It is
true that these high-symmetry points usually reflect the extrema in the
PES. However, due to this limitation the fitted continuous PES can only 
contain terms that correspond to these high-symmetry situations. 
On the one hand this often saves computer time also in the quantum
dynamics because certain additional selection rules are introduced 
which reduces the necessary basis set \cite{Wie96,Gro95PRL}.
On the other hand, of course this represents an approximation.
The question, how serious the neglect of terms with lower symmetry is, 
remains open until these terms have been determined and included in 
actual dynamical calculations.

If more than just the molecular degrees of freedom should be considered
in a parametrization of an {\it ab initio} PES, analytical forms become 
very complicated and cumbersome. As an alternative, the interpolation
of {\it ab initio} points by a neural network has been 
proposed \cite{Bla95,No97,Lor98}. 
Neural networks can fit, in principle, any real-valued,
continuous function to any desired accuracy. They 
require no assumptions of the functional form of the underlying problem.  
On the other hand, there is no physical insight that is used as an input
in this parametrization. Hence the parameters of the neural network do
not reflect any physical or chemical property. Another approach
is to fit the parameters of a tight-binding formalism to 
{\it ab initio} energies \cite{Mehl96}. A tight-binding method
is more time-consuming than an analytical representation or 
a neural network since it requires the diagonalization of a matrix.
However, due to the fact that the quantum mechanical nature of bonding is 
taken into account \cite{Gor97} tight-binding schemes
need a smaller number of {\it ab initio} input points to perform a good
interpolation {\em and} extrapolation \cite{Gro98TB}. In addition,
their parameters, the Slater-Koster integrals \cite{Sla54}, have
a physical meaning.
It remains to be seen how useful the different fitting methods will be for 
performing {\it ab initio} dynamics simulations.

It has been argued that any fitting scheme using analytic potential
functions introduces a possible bias into the parametrization 
\cite{Rad97Rev}. In order to avoid this one should use {\it ab initio}
molecular dynamics methods that calculate the forces ``on the fly''
because this requires no fitting. However, approaching a problem
apparently unbiased also means not to take advantage of any previous
knowledge. On the contrary, using as much experience and knowledge
as possible should usually make any method more efficient \cite{Mak98}.

An important issue is to judge the quality of the fit to an 
{\it ab initio} PES. Usually the root mean squared (RMS) error between fit and
input data is used as a measure of the quality of a fit. If this
error is zero, then everything is fine. However, normally this
error is larger than zero. The systematic error of the {\it ab initio} 
energies is usually estimated to be of the order of 0.1~eV. Often it is said 
that the RMS error of the fit should be of the same order.
But the dynamics of molecular dissociation at surfaces can be dramatically
different depending on whether there is a barrier for dissociation
of height 0.1~eV or not \cite{Gro98PRL}. Hence for certain regions of the
PES the error has to be much less than 0.1~eV, while for other regions
even an error of 0.5~eV might not influence the dynamics significantly.
Another example occurs in a reaction path parametrization. If the curvature
in the parametrization is off by a few percent, the energetic distribution of
barrier heights is not changed and the dynamical properties
are usually not altered significantly. However, the location of the barriers is
changed and consequently the RMS error can become
rather large. Hence one has to be cautious by just using the RMS error
as a quality check of the fit. Unfortunately there is no other simple
error function for the assessment of the quality of a fit. 
If it is possible, one should perform a dynamical check. Obviously,
if the dynamical properties calculated on a fitted PES agree with the
ones calculated on the original PES, the quality of the fit should be 
sufficient.

\subsection{Quantum dynamics}
There are two ways to determine quantum mechanical reaction probabilities:
by solving the time-dependent or the time-independent Schr{\"o}dinger
equation. Both approaches are equivalent \cite{New82} and should give the
same results. The chosen method depends on its applicability,
but apparently it is often also a matter of training and personal taste.

In the most common time-independent formulation, the concept of defining one
specific reaction path coordinate is crucial. Starting from the 
Schr{\"o}dinger equation
\begin{equation}
(H \ - \ E) \ \Psi \ = \ 0,
\label{TISE}
\end{equation}
one chooses one specific reaction path coordinate $s$ and separates
the kinetic energy operator in this coordinate
\begin{equation}
(\frac{-\hbar^2}{2 \mu} \partial_s^2 \ + \ \tilde H \ - E) \ \Psi \ = \ 0.
\label{ReacHam}
\end{equation}
Here $\tilde H$ is the original Hamiltonian except for the kinetic energy
operator in the reaction path coordinate.
Usually the use of curvelinear reaction path coordinates results in a more
complicated expression for the kinetic energy operator involving cross terms
\cite{Hof63,Mar64,Bre89}, but for the sake of clarity I have neglected this
in Eq.~\ref{ReacHam}. As the next step one expands the wave function in the
coordinates perpendicular to the reaction path coordinate in some suitable
set of basis functions,  
\begin{equation}
\Psi \ = \Psi (s,\ldots) \ = \ \sum_n \ \psi_n(s) \ | n \rangle.
\label{expa}
\end{equation} 
Here $n$ is a multi-index, and the expansion coefficients $\psi_n(s)$
are assumed to be a function of the reaction path coordinate.
Now we insert the expansion of $\Psi$ in Eq.~\ref{ReacHam} and 
multiply the Schr{\"o}dinger equation by $\langle m |$, which
corresponds to performing a multi-dimensional integral. Since the 
basis functions $| n \rangle$ are assumed to be independent of $s$,
we end up with the so-called coupled-channel equations,
\begin{equation}
\sum_n \left\{ (\frac{-\hbar^2}{2 \mu} \partial_s^2 \ - \ E ) \ \delta_{m,n}
+ \ \langle m | \tilde H | n \rangle \right\}  \ \psi_n (s) \ = \ 0.
\label{CC}
\end{equation} 
Instead of a partial differential equation -- the original time-independent
Schr{\"o}dinger equation Eq.~\ref{TISE} -- we now have a set of coupled
ordinary differential equation. Still a straightforward numerical integration
of the coupled-channel equations leads to instabilities, except for in 
simple cases, due to exponentially increasing so-called closed channels.
Recently a very stable and efficient coupled-channel algorithm has been
introduced \cite{Bre93,Chi94,Bre94,Bre95}. The main idea underlying this
particular algorithm  will be briefly sketched in the following.

For the solution $\Psi$ defined in Eq.~\ref{expa}, which represents a vector 
in the space of the basis functions, the initial conditions
are not specified. This function can also be considered as a matrix
\begin{equation}
\Psi \ = \ (\psi)_{nl},
\label{mat}
\end{equation}
where the index~$l$ labels a solution of the Schr{\"o}dinger equation
with an incident plane wave of amplitude one in channel~$l$ and zero
in all other channels. Formally one can then write the
solution of the Schr{\"o}dinger equation for a scattering problem
in a matrix notation as
\begin{eqnarray}
\Psi (s \rightarrow + \infty) & = & e^{-iqs} \ - \ e^{iqs} \ r, \nonumber\\
\Psi (s \rightarrow - \infty) & = & e^{-iqs} \  t.
\end{eqnarray}
Here $q = q_m \delta_{m,n}$ is a diagonal matrix, $r$ and $t$ are the
reflection and transmission matrix, respectively. Now one makes the
following ansatz for the wave function,
\begin{equation}
\Psi (s) \ = \ \left( 1 - \rho(s) \right) \ \frac{1}{\tau(s)} \ t.
\label{LI}
\end{equation} 
Equation~\ref{LI} defines the {\em local reflection matrix}~$\rho(s)$
($LORE$) and the {\em inverse local transmission matrix}~$\tau(s)$
($INTRA$). The boundary values for these matrices are (except for
phase factors which, however, do not affect the transition probabilities):
\begin{equation} 
\left( \rho(s); \tau (s) \right) \ = \ \left\{ 
\begin{array}{ll}
(r;t) & s \rightarrow + \infty \\
(0;1) & s \rightarrow - \infty 
\end{array} \right.
\end{equation}
From the Schr{\"o}dinger equation first order differential equations
for both matrices can be derived \cite{Bre93,Chi94,Bre94,Bre95} which can
be solved by starting from the known initial values at 
$s \rightarrow - \infty$; at $s \rightarrow + \infty$ one then obtains 
the physical reflection and transmission matrices. Thus the numerically 
unstable boundary value problem has been transformed to a stable initial 
value problem.

In the time-dependent or wave-packet formulation, 
the solution of the time-dependent
Schr{\"o}dinger equation 
\begin{equation}
i\hbar \ \frac{\partial}{\partial t} \ \Psi({\bf R},t) \ =
\ H \ \Psi({\bf R},t) 
\end{equation}
can formally be written as
\begin{equation}
\Psi({\bf R},t) \ = \ e^{-i H t/\hbar} \ \Psi({\bf R},t=0),
\end{equation}
if the potential is time-independent. The most common methods to
represent the time-evolution operator $\exp(-i H t/\hbar)$ 
in the gas-surface dynamics community are the
split-operator \cite{Fle76,Feit82} and the Chebychev \cite{Tal84} methods.
In the split-operator method, the time-evolution operator for small
time steps $\Delta t$ is written as
\begin{equation}
e^{-i H \Delta t/\hbar} \ = \ e^{-i K \Delta t/2\hbar} \ 
e^{-i V \Delta t/\hbar} \ e^{-i K \Delta t/2\hbar} \ + \ O(\Delta t^3),
\end{equation}
where $K$ is the kinetic energy operator and $V$ the potential term.
In the Chebyshev method, the time-evolution operator is expanded as
\begin{equation}
e^{-i H \Delta t/\hbar} \ = \ \sum_{j=1}^{j_{\rm max}} \ a_j(\Delta t) \
                              T_j (\bar{H}),
\end{equation}
where the $T_j$ are Chebyshev polynomials and $\bar{H}$ is the Hamiltonian
rescaled to have eigenvalues in the \mbox{range (-1,1)}. Both propagation 
schemes use the fact that the kinetic energy operator is diagonal in
${\bf k}$-space and the potential in real-space. The wave function and
the potential are represented on a numerical grid, and the 
switching between the ${\bf k}$-space and real-space representations is 
efficiently done by Fast Fourier Transformations (FFT) \cite{NumRec}.

In the last years the time-dependent wave-packet methods have been much more
fashionable than time-independent schemes in the gas-surface dynamics
community. Many different groups have used wave-packet codes to study
the dissociation dynamics on surfaces, up to recently almost entirely
on low-dimensional model potentials \cite{Jac87,Hal90,Dar92a,Han90,%
Mow93,Dar94JCP,Dai95,Mue87,Lun91,Kra91}. It has been argued that wave-packet
methods avoid "the problem of excessively many channels" \cite{Dar95rep}. 
But actually these methods suffer from the fact that the wave function
and the potential are represented on a grid which leads to memory problems
in the implementation. In each dimension between 16 and several hundred
grid-points are used. In time-independent methods, on the contrary, it is
for example sufficient to expand the hydrogen wave function in the interatomic
distance  in three or four eigenfunctions of a harmonic oscillator
due to the large vibrational energy quantum. One might say that the hydrogen
vibration in time-independent methods is represented by three or four 
points instead of more than sixteen in time-dependent methods.
In addition, while in the $LORE-INTRA$ scheme \cite{Bre93,Chi94,Bre94,Bre95}
the matrices are successively determined along the reaction path and do not 
have to be stored, in wave-packet methods the wave function has always
to be stored everywhere on the grid. 
The saving in storage requirements is one of the reasons that the first
six-dimensional quantum dynamical treatment of hydrogen dissociation on
surfaces was indeed a time-independent one using the $LORE-INTRA$ scheme
\cite{Gro95PRL}.

\subsection{Classical dynamics}

Once an analytical continuous representation of an {\it ab initio} PES is 
available, it is also possible to determine the gradients of the potential
analytically. This allows one to perform {\it ab initio} molecular dynamics
studies by integrating the classical equations of motion,
\begin{equation}
M_i \ \frac{\partial^2}{\partial t^2} {\bf R}_i \ = \
- \frac{\partial}{\partial {\bf R}_i} V (\{ {\bf R}_j \} ).
\end{equation}
The solution of the equations of motion can be obtained by standard
numerical integration schemes like Runge-Kutta, Burlisch-Stoer or
predictor-corrector methods (see, e.g., Ref. \cite{NumRec}). Sticking
corresponds to a process in which statistically distributed particles
hit the surface. This means that the determination of classical
sticking probabilities requires an average over typically thousands of
trajectories. The initial conditions can be chosen either by some
Monte-Carlo sampling or by an equidistant sampling.

Most {\it ab initio} total-energy programmes based on plane-wave
expansions also evaluate the gradients of the potential via the
Hellmann-Feynman theorem\cite{Hell37,Feyn39}. With these forces the classical
equations of motion can directly be solved. This ``traditional''
{\it ab initio} molecular dynamics scheme with the determination
of the forces on the fly has the advantage that it does not
require the fitting of the {\it ab initio} PES to any analytical
or numerical representation. This is especially advantageous if
surface degrees of freedom play an important role in the
dissociation process. On the other hand, every step of the
numerical integration of the equations of motion requires a new
{\it ab initio} total-energy calculation which is still rather
time-consuming. Hence the number of trajectories obtainable 
in such a ``during the journey'' {\it ab initio}
molecular dynamics simulation is limited to numbers well 
below 100 \cite{DeVita93,Gro97H2Si,Silva97,Rad97Rev}. Such a simulation only 
makes sense if one can extract meaningful information from a low number
of events. For the simulation of adsorption events this is usually
not the case, but for desorption, where the particles originate
from a small portion of the phase space, ``on the fly'' 
{\it ab initio} molecular dynamics simulations can still be
useful \cite{Gro97H2Si}.

But even if an analytical representation of the {\it ab initio} PES
is available, there is still some computational effort to determine
dissociation probabilities. It is a wide-spread believe that classical 
dynamical methods are much less time-consuming than
quantum ones. This is certainly true if one compares the computational
cost of one trajectory to a quantum calculation. 
The quantum calculations, however, take advantage of the delocalized
nature of the wave functions. A beam of particles approaching the surface
is described by a plane wave in the gas phase that includes all impact points.
And a non-rotating molecule is represented by an $j = 0$ rotational
state, where $j$ is the rotational quantum number and which contains
all orientations with equal probability. Hence the averaging over 
initial conditions is done automatically in quantum mechanics.
Consequently, quantum dynamical simulations do not necessarily
have to be more time-consuming that classical calculations, 
in particular if one considers the fact that in wave-packet calculations
the dissociation probability for a whole range of energies is
calculated in one run or that in a time-independent coupled-channel method the
sticking and scattering probabilites of all open channels are
determined simultaneously.

The crucial difference between quantum and classical dynamics is
that in the quantum dynamics the averaging is done coherently while it is
done incoherently in the classical dynamics. For heavier molecules
this is apparently not important, but for hydrogen dissociation
this has important consequences, as will be shown below.

\section{Hydrogen dissociation on metal surfaces}

After introducing the theoretical concepts necessary to perform
{\it ab initio} dynamics calculations, I will now discuss applications.
I start with the hydrogen dissociation on metal surfaces.

While the dissociation of hydrogen at simple or noble metals is usually 
hindered by a substantial energy barrier, transition metal surfaces
are rather reactive, i.e., already at low kinetic energies the
dissociative adsorption probability is often larger than 10 per cent. 
This reactivity has been attributed to the local density of states (LDOS)
of the partly filled $d$-band at the Fermi level \cite{Feib84,Har85}
or close to the Fermi level \cite{Wil96PRL,Wil96AP} of the transition metals.
Recent investigations have shown that the reactivity cannot be solely
understood in terms of the LDOS at the Fermi level, but rather by the
hybridization between molecular and metal states which involves the whole 
$d$-band \cite{Ham95PRL,Ham95Nat,Ham95SS}. 

In noble metals the $d$-band is filled, which leads to the existence
of a significant barrier towards dissociative adsorption of hydrogen. 
First I will discuss the {\it ab initio} dynamics calculations for such 
an activated system, namely H$_2$/Cu which has been the benchmark system for 
the study of dissociative adsorption. Then I will address the dissociation
on transition metals surfaces. For these latter systems {\it ab initio} 
dynamics calculations have proven their power by demonstrating the hitherto 
underestimated efficiency of one particular dynamical mechanism, 
the steering. This chaper ends with a discussion of the influence of
preadsorbates on the energetics and dynamics of hydrogen dissociation
on metal surfaces.

\subsection{Dissociation on simple or noble metals}

{\em The} model system for the study of dissociative adsorption over the
last years has been H$_2$/Cu. Many detailed 
experimental \cite{Ret92,Ret95,Hay89,Ang89,Ren94} as well as theoretical 
\cite{Dar92a,Kue91,Mow93,Dar94JCP,Bru94,Kum94,Dai95,Nie90,Eng92,Eng93}
dynamical investigations about this system exist. The early dynamical 
calculations were mostly performed on model potentials. Sometimes the
used potential energy surfaces were  derived from a small
cluster calculation where the Cu surface had been modelled by a Cu$_2$ 
dimer \cite{Har85}, or the PES was obtained from approximate methods like
the effective medium theory \cite{Nie90,Eng92,Eng93}.

\begin{figure}[tb]
\unitlength1cm
\begin{center}
   \begin{picture}(10,8.0)
\put(-1.,1.){ {\epsfxsize=10.cm  
          \epsffile{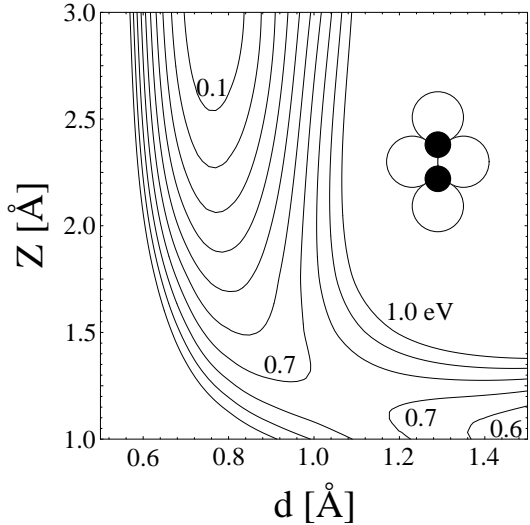}} }

   \end{picture}
\end{center}

\vspace{-1.cm}

   \caption{Contour plot of the PES along a two-dimensional cut 
            through the six-dimensional coordinate space of 
            H$_2$/Cu(111) determined by DFT-GGA calculations. 
            The inset illustrates the
            orientation of the molecular axis and the lateral
            H$_2$ center-of-mass coordinates, i.e. the coordinates
            $X$, $Y$, $\theta$, and $\phi$. The coordinates 
            in the figure are the H$_2$ center-of-mass distance 
            from the surface $Z$ and the H-H interatomic distance $d$. 
            Energies are in eV per H$_2$ molecule.
            The contour spacing is 0.1~eV. This
            cut corresponds to the minimum energy
            pathway (from ref.~\protect\cite{Ham94}).}

\label{H2Cuelbow}
\end{figure}

Then in 1994 there appeared two papers back-to-back in Physical Review Letters
\cite{Ham94,Whi94} which represented the first detailed determination of a 
multi-dimensional PES as a function of all six hydrogen degrees of freedom
by density-functional calculations. First of all these calculations showed 
that one has to go beyond the local density approximation (LDA) for the
treatment of the exchange-correlation effects in density-functional
theory calculations. LDA calculations yield an almost vanishing barrier 
to dissociation although the experiments show that the height of this barrier 
should be about 0.5~eV \cite{Ret92,Ret95,Hay89,Ang89,Ren94}. By including 
gradient corrections in the exchange-correlation functional 
(the so-called generalized gradient approximation (GGA) \cite{Per92}) the 
general agreement with the experiment is greatly improved. 
Figure \ref{H2Cuelbow} shows the GGA minimum barrier for 
H$_2$/Cu(111) in a so-called elbow plot. The PES is plotted as a function 
of the H$_2$ center-of-mass distance from the surface and the interatomic 
H-H distance. The molecular orientation and the center-of-mass coordinates 
parallel to the surface are kept fixed. The calculations to determine the 
PES are not trivial; care has to be 
taken that the results are well-converged \cite{Ham94,Kra96HCu}, 
and there are still some uncertainties about the exact barrier 
heights \cite{Whi94,Kra96HCu,Wie95}. GGA represents a great improvement 
over LDA for certain systems as far as barrier heights are concerned, 
but it has its limitations \cite{Per92,Nac96}, and the development of more 
accurate functionals is still going on (see, e.g., Ref.~\cite{Per96,Staed97}).

\begin{figure}[tb]
\unitlength1cm
\begin{center}
   \begin{picture}(10,8.0)
{ {\epsfxsize=6.cm  
          \epsffile{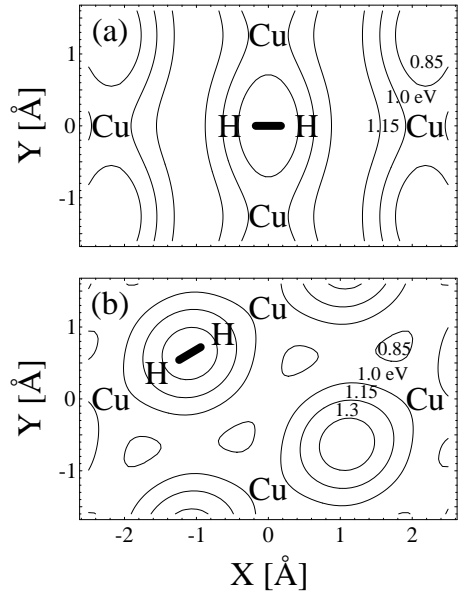}} }

   \end{picture}
\end{center}

\vspace{-0.5cm}

   \caption{The dissociation barrier height for H$_2$/Cu(111)
            as a function of the two lateral H$_2$ 
            center-of-mass coordinates $X$ and $Y$ 
            determined by DFT-GGA calculations. The
            positions of the copper substrate atoms are
            marked by ``Cu''. The molecular axis is parallel
            to the surface, its azimuthal orientation is
            illustrated by ``H-H'' (from ref.~\protect\cite{Ham94}).}     

\label{H2Cucorr}
\end{figure}

The two papers \cite{Ham94,Whi94} represented indeed a breakthrough 
in the determination of accurate and detailed potential energy surfaces. 
They challenged the gas-surface theorists with the fact that the interaction
of hydrogen molecules with metal surfaces was much more corrugated than
was anticipated before. Since the electronic density in front of metal
surfaces is rather smeared out \cite{Feib84,Che75}, it was assumed that
the barrier for dissociation depends only very weakly on the location
within the surface unit cell. This picture of a flat, structureless
surface was seemingly confirmed by the experimentally found normal energy 
scaling of the sticking probability \cite{Ret92,Ren94,Mic91}, i.e., the
sticking probability is a function of the normal component of the
incident kinetic energy alone.

The {\it ab initio} calculations, however, showed that the barrier to 
dissociative adsorption for the molecular axis parallel to the surface varies 
by 0.7~eV within the surface unit cell (see Fig.~\ref{H2Cucorr}) \cite{Ham94}.
The bond-breaking process even on metal surfaces is a very localized process
close to the surface. It involves the hybridization of molecular orbitals
with particular metal states, in particular $d$-states. These have 
a spatially strongly varying distribution reflecting their symmetry properties.
For an illustration of the local character of this interaction see for example
Refs. \cite{Wil96AP,Eich98}. Hence it is the chemical nature of the 
dissociation process that leads to its spatial variation.

Motivated by the results of the {\it ab initio} calculations
for the H$_2$/Cu PES, Darling and Holloway showed by using a three-dimensional
model PES that strong corrugation can be reconciled with normal energy
scaling \cite{Dar94}. For this to happen, the higher barriers have to be
further away from the surface than the lower ones. This ``balanced''
corrugation \cite{Dar95rep} was indeed found in the {\it ab initio}
calculations \cite{Ham94,Whi94}.  

Shortly after the publication of the {\it ab initio} H$_2$/Cu(111) PES
the first high-dimensional dynamical calculations were performed 
\cite{Gro94PRL} where the PES was entirely based on the 
first-principles calculations \cite{Ham94}. A five-dimensional
parametrisation of the {\em ab initio} PES has served as an input
for the time-independent coupled-channel study.
In this study the three center-of-mass coordinates and the interatomic
spacing of the H$_2$ molecule have been treated quantum mechanically, while
the azimuthal orientation of the molecule has been taken into account in a
classical sudden approximation which works quite well in the H$_2$/Cu
system \cite{Gru93}. 

\begin{figure}[tb]
\unitlength1cm
\begin{center}
   \begin{picture}(10,6.5)
      \includegraphics{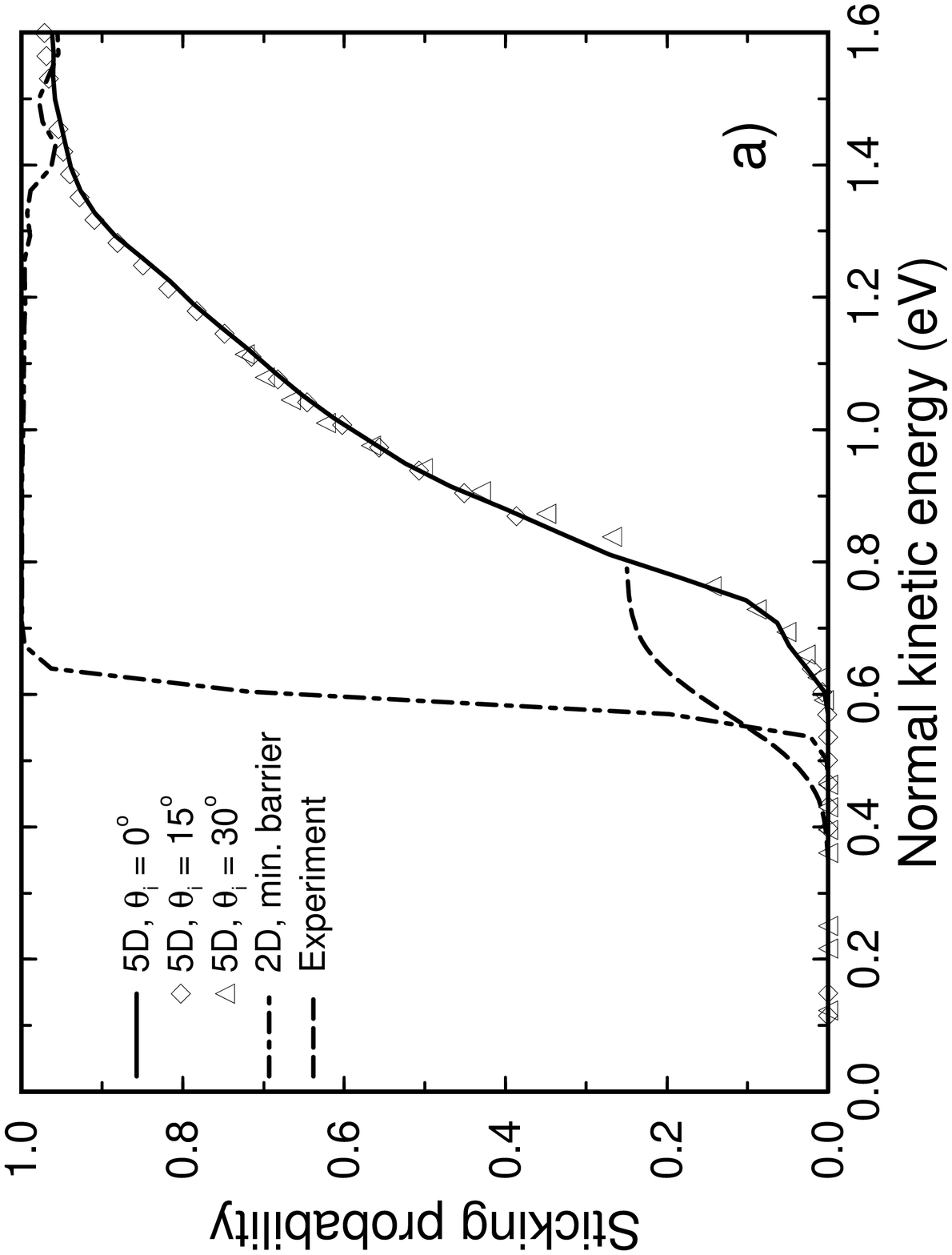}
   \end{picture}
   \begin{picture}(10,6)
      \includegraphics{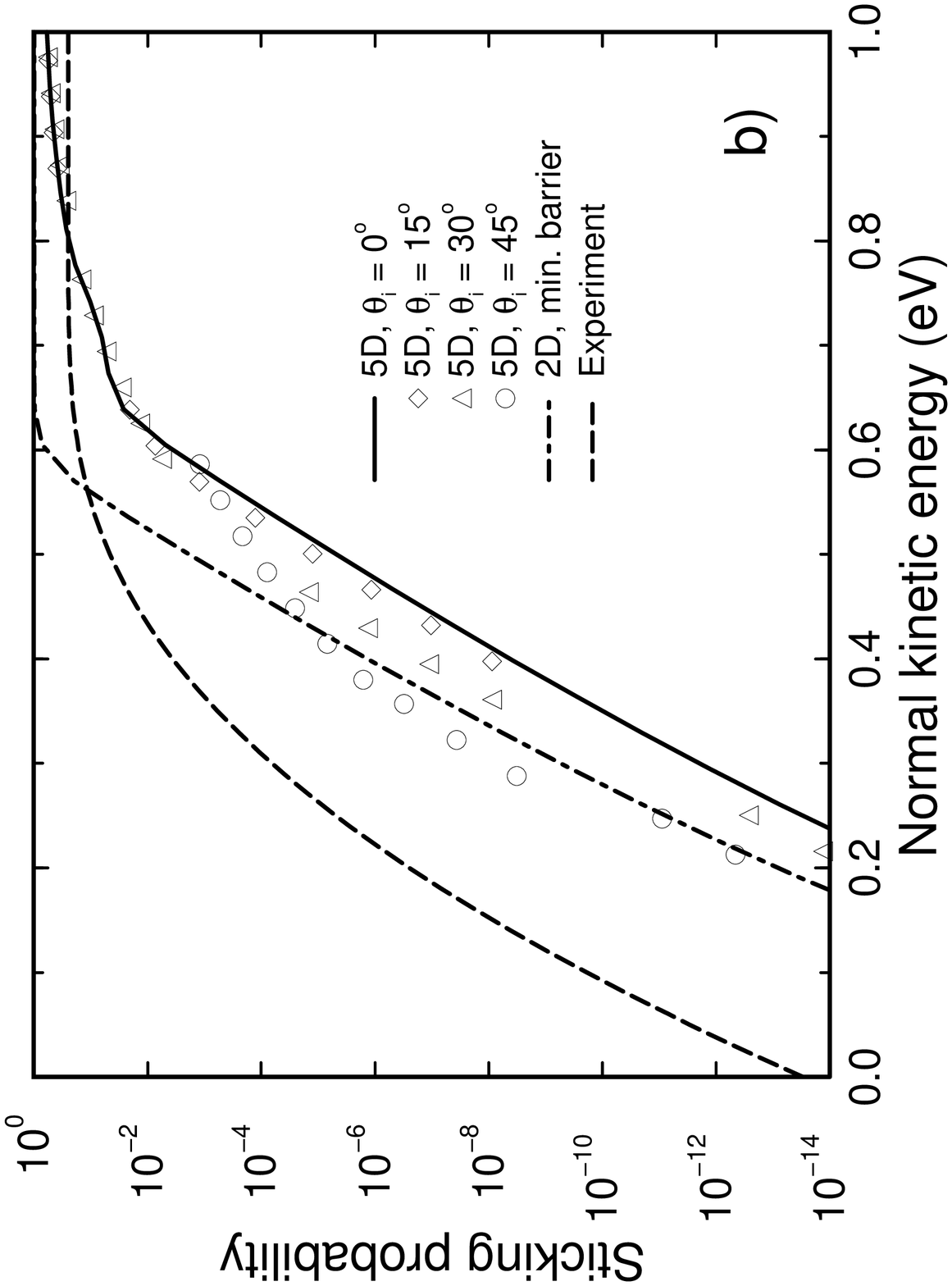}
   \end{picture}
\end{center}
   \caption{Sticking probability versus normal kinetic energy of H$_2$/Cu(111)
        for molecules initially in the vibrational ground state. 
a) Linear plot, b) logarithmic plot (note the different energy range). 
5D-calculations for different incident angles at the corrugated
surface: solid line $\theta_i = 0^{\circ}$, 
$\Diamond \ \theta_i = 15^{\circ}$, $\bigtriangleup \ \theta_i = 30^{\circ}$,
$\bigcirc \ \theta_i = 45^{\circ}$;
Dash-dotted line: 2D-calculations corresponding to a flat surface 
with the minimum barrier (from ref.~\protect\cite{Gro94PRL}).
Dashed line: Experimental curve (from ref.~\protect\cite{Ret95}).} 
\label{stickh2cu}
\end{figure}

These calculations demonstrated the importance of the multi-dimensionality for
the determination of the sticking probability. Before, in the low-dimensional
studies the barrier thickness has been used as a variable parameter in order
to reproduce the slope of the sticking curve. In low-dimensional calculations 
the width of the increase is basically caused by tunneling for energies below 
the barrier height and by quantum back scattering for energies above the 
barrier height. Furthermore, mixed quantum-classical calculations of the 
sticking probability for H$_2$/Cu obtained by effective medium theory 
\cite{Eng92,Eng93} showed only a modest dependence of the width of the 
sticking curve on the dimensionality of the calculations. The 5D calculations 
on the GGA-PES demonstrated that only in the tunneling regime the rise of the 
sticking probability is determined by the barrier thickness \cite{Gro94PRL}. 
Figure~\ref{stickh2cu} shows the results for the calculated sticking
probabilities of these 5D calculations. They are also compared to 2D 
calculations that included only the minimum energy path geometry plotted
in Fig.~\ref{H2Cuelbow}. For energies above the minimum barrier 
height the results of the 2D and the 5D calculations are very different. 
In this regime it is the whole barrier distribution that determines the 
width of the sticking curve. In the tunneling regime, on the other hand,
the sticking curves of the 2D and the 5D calculations are just shifted
(Fig.~\ref{stickh2cu}b) demonstrating that in this regime it is indeed
the minimum barrier thickness that governs the adsorption dynamics.

As mentioned above, in the low-dimensional studies it was basically
the barrier width that determined the rise of the sticking probability
with kinetic energy. To reproduce the gradual increase of the
experimental results, small barrier widths had to be assumed which
were much smaller than the results obtain in the GGA calculations.
In addition, to account for the experimental fact that vibrational excitation
of the impinging molecules is very efficient for increasing the sticking
probability \cite{Ret92}, a late barrier after a strongly curved region
according to the Polanyi's rules \cite{Pol69} had been assumed in
the model PES \cite{Dar92a,Kue91,Han90}. The curvature of the
reaction path in the GGA-PES (Fig.~\ref{H2Cuelbow}) is much smaller
than was previously assumed. It is not only the curvature, but also
the lowering of the vibrational frequency in the barrier region, which
on the other hand was underestimated before, that
causes the vibrational effects in the dissociation process \cite{Gro96CPLa}.

With regard to the effect of additional parallel momentum,
Fig.~\ref{stickh2cu} shows that for non-normal incidence 
the sticking probabilities fall upon the normal incidence sticking
curve if they are plotted versus the normal kinetic energy
for kinetic energies larger than 0.6~eV, i.e. in the regime where
sticking is classically possible. 
The GGA-PES indeed shows the topological features necessary
for normal energy scaling \cite{Dar94}, namely the balanced
corrugation \cite{Dar95rep} with the larger barriers further
away form the surface than the smaller. The 5D-calculations
thus confirm the 3D model calculations \cite{Dar94} with
respect to the role of the surface corrugation.

However, for the tunneling regime Fig.~\ref{stickh2cu}b) reveals that 
additional parallel momentum enhances sticking. This is seemingly at 
variance with the molecular beam adsorption experiments in which 
normal energy scaling has been found for all energies in the system 
H$_2$/Cu(111) \cite{Ret92,Ret95}. However, this apparent discrepancy
can be explained by the fact that at low kinetic energies sticking in the beam
experiments is dominated by the vibrationally excited molecules, and for
these molecules the range where normal-energy scaling is obeyed is shifted
to lower energies \cite{Bre95}. In very recent experiments \cite{Mur98}
the state resolved angular dependence of the sticking probability of
H$_2$/Cu(111) has been determined by applying detailed balance
arguments to desorption experiments. And indeed, these experiments
confirmed the theoretical prediction that additional parallel momentum
enhances the sticking probability in the tunneling regime.
In this tunneling regime the classical conception of particles moving
like bob-sleds in a corrugated potential does not apply any more.
The shortest propagation path through the barrier region is exponentially
preferred with respect to all other paths. This shortest path usually 
corresponds to normal propagation through the barrier for a corrugated
potential. Due to the corrugation parallel momentum can be efficiently 
transferred to normal momentum for tunneling particles, and this causes 
the enhancement of the sticking probability for additional parallel momentum
in the low-energy regime.

\begin{figure}[tb]
\unitlength1cm
\begin{center}
   \begin{picture}(10,8.0)
\put(1.,-1.){ {\epsfxsize=7.5cm  
          \epsffile{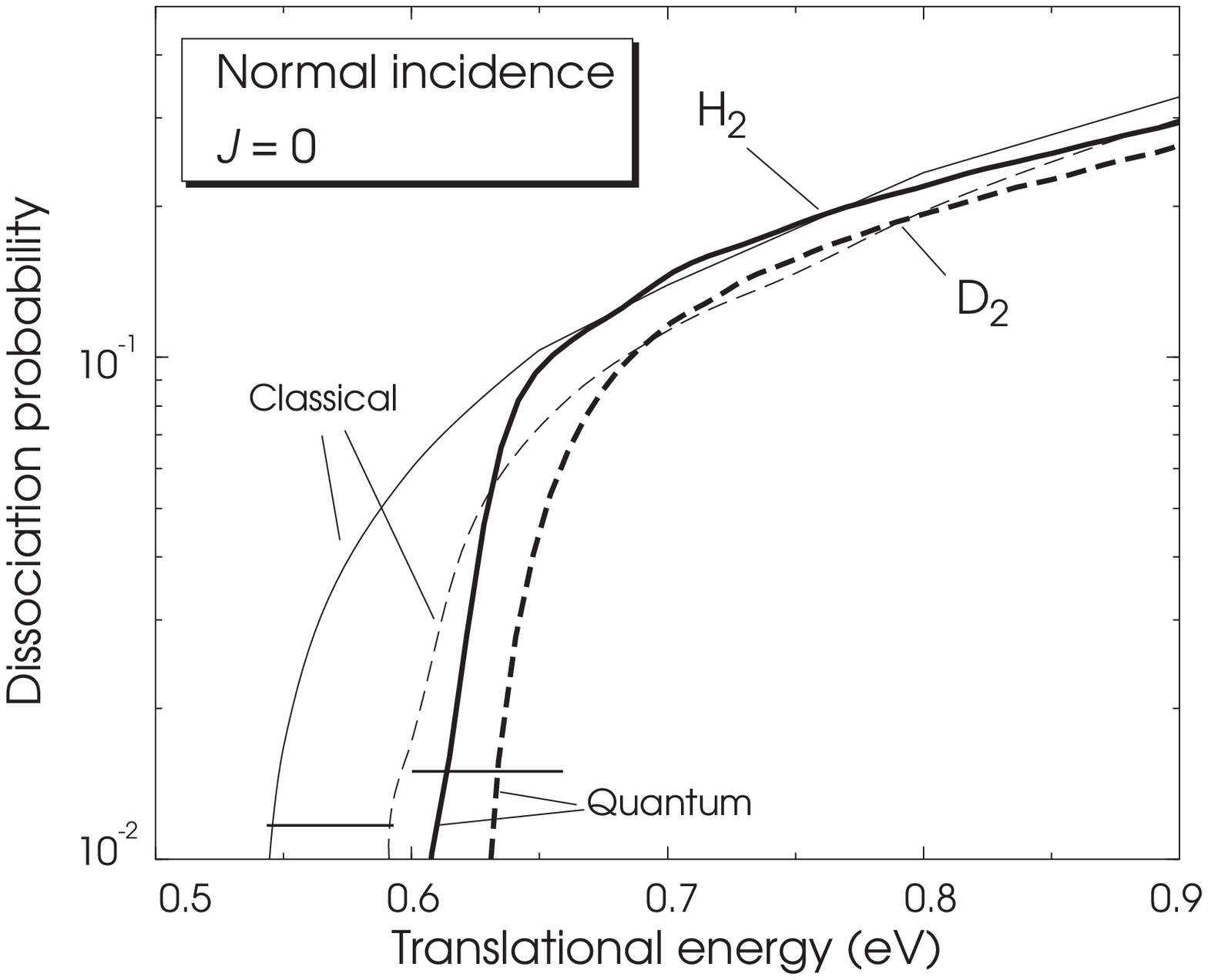}} }

   \end{picture}
\end{center}

\vspace{-1.cm}

   \caption{Classical and quantum dissociation probability as a
            function of kinetic energy for initially non-rotating
            H$_2$ and D$_2$ molecules under normal incidence at Cu(111),
            determined by three-dimensional wave-packet 
            calculations \protect\cite{Kin96}. In these calculations
            the center-of-mass distance from the surface, one 
            center-of-mass coordinate parallel to the surface and the
            polar orientation of the molecule were taken into account.}

\label{qmclass}
\end{figure}

The GGA-PES for H$_2$/Cu(111) of Hammer and coworkers served also
as an input for three-dimensional wave-packet calculations \cite{Kin96} 
in which, besides the H$_2$ center-of-mass distance from the surface, 
one H$_2$ center-of-mass coordinate parallel to the surface and the
polar orientation of the molecule were considered. The change of the
vibrational frequency was taken into account by keeping the molecule
adiabatically in its vibrational ground state. This study particularly
focused on the difference between the results of quantum and classical
calculations. Figure~\ref{qmclass} shows the dissociation probability
as a function of the kinetic energy for energies close to the minimum
barrier height. Classical and quantum results are shown for the two 
hydrogen isotopes H$_2$ and D$_2$. The difference between the two isotopes
is due to the decrease in the zero-point vibrational energy during
the dissociation process. Since the vibrational frequency of D$_2$ is
smaller by a factor of 1/$\sqrt{2}$ than the H$_2$ frequency, for D$_2$
less vibrational energy is transferred to the translational degree of
freedom leading to the smaller dissociation probability.

Rather surprising is the fact that for energies
close to the dissociation barrier 
$E \approx 0.65$~eV the classical probabilities are larger than the
quantum ones. Naively one would expect that the quantum results should
be larger due to tunneling which is absent in the classical calculations.
However, the PES is corrugated and anisotropic. For energies close to the 
minimum barrier energy only a limited range of molecular orientations
and lateral coordinates are energetically accessible at the barrier
position. The localisation of the wave function in the angular and
lateral degrees of freedom during the crossing of the minimum barrier 
region leads to the building up of zero-point energies in the quantum
calculations. These zero-point energies cause an effective increase
of the barrier height in the quantum dynamics which reduces the quantum
dissociation probability compared to the classical probability. The
zero-point energies also cause quantization effects in the dissociation
probability at higher energies which are visible as a weak step-like
structure in Fig.~\ref{qmclass}.

\begin{figure}[tb]
\unitlength1cm
\begin{center}
   \begin{picture}(10,7.0)
\put(-.5,-5.5){   {\epsfxsize=10.cm  
          \epsffile{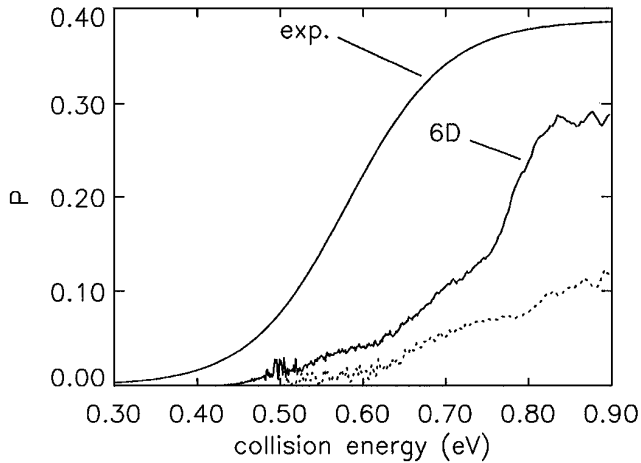}} }
   \end{picture}
\end{center}
\caption{Sticking probability for H$_2$ dissociation on Cu(100) 
determined by six-dimensional time-dependent quantum calculations 
on an {\it ab initio} PES (from Ref.~\protect\cite{Kro97PRL}). 
The experimental data are taken 
from Ref.~\protect\cite{Mic91}. The dotted line shows the calculated
probability for vibrational excitation in scattering.}
\label{HCu6D}
 
\end{figure}

Just recently the first six-dimensional wave-packet studies of the dissociation
of H$_2$ on Cu have appeared \cite{Kro97PRL} which confirmed the
importance of the multi-dimensionality for the dissociation process.
Fig.~\ref{HCu6D} shows the results for the sticking probability 
of H$_2$ on Cu(100) \cite{Kro97PRL}. The calculations were based
on a analytical fit to the PES obtained by density functional theory
using  GGA. The wave-packet method used a symmetry-adapted basis 
set in the center-of-mass degrees of freedom parallel to the surface and
for the molecular orientation. This causes a significant computational
saving for initial conditions corresponding to normal incidence. 6D
wave-packet calculations for non-normal incidence have not been
performed yet. The theoretical results show some structure, in particular
close to the threshold at 0.4~eV. These oscillations are attributed to
resonances caused by the weakening of the molecular bond close to
the surface \cite{Kro97PRL}.

The 6D sticking curve seems to be shifted by roughly 0.2~eV to higher 
energies with respect to the experimental results. A similiar shift is 
found in the 5D calculations for H$_2$/Cu(111) in Fig.~\ref{stickh2cu}.
This suggests that the GGA calculations overestimate the dissociation
barriers. For H$_2$/Cu(111), convergence tests of the {\it ab initio}
calculations suggested that the barriers should be lowered by 0.2~eV
\cite{Ham94}; the inclusion of the polar orientation, which was not
considered in the 5D calculations, would shift the sticking curves to
higher energies again, though.

On the other hand, the sticking probability is directly measured only
for kinetic energies up to approximately 0.5~eV. H$_2$ beams with
higher kinetic energies can not be prepared in nozzle experiments.
The experimental sticking probabilty for higher kinetic energies is
derived from desorption experiments invoking the principle of
detailed balance or microscopic reversibility. Because of this
indirect procedure, for example the saturation value of the sticking
probability, i.e. the maximum value at high energies, is not
well-established experimentally; it has an uncertainty of a factor
of two \cite{Ret95}. Hence it is not entirely clear yet, whether the
discrepancy between theory and experiment
is caused by experimental uncertainties or by approximations
in the calculations.

Another six-dimensional time-dependent dynamical study is devoted
to the dissociation of H$_2$ on Cu(111) \cite{Dai97}. However, this
study is not fully based on an {\it ab initio} PES. The authors claim
that the available {\it ab initio} information is not complete enough,
however, they have not collaborated with any group that is able
to perform {\it ab initio} total-energy calculations.
They use a LEPS potential for the parametrization of the H$_2$
interaction with Cu(111). The LEPS parametrization is considered
by other authors \cite{Wie96} to be not flexible enough to accurately 
describe the PES.  

Although already many detailed studies exist about the system
H$_2$/Cu, experimentally as well as theoretically, there is still
room for further explorations in this system. For example, current
experimental studies have focused on the rotational alignment
of desorbing molecules \cite{Wet96Euro,Gul96,Hou97}. These investigations
address the issue of the anisotropy of the dissociation barriers. 
Another noble metal system under current interest is the dissociation 
of hydrogen on silver. Detailed experimental data exist for this 
system \cite{Cot97,Mur97}, and just recently also the PES for H$_2$/Ag(100)
was determined by first-principles calculations \cite{Eich98}. 
Hence there are still interesting results to be anticipated for the 
interaction of hydrogen with noble or simple metal surfaces.

\subsection{Dissociation on transition metals}

Molecular beam experiments of the dissociative adsorption of H$_2$ on Pd(100) 
have found that the sticking probability initially decreases with increasing 
kinetic energy of the beam \cite{Ren89,Ret96}. Similiar results have 
been obtained for the H$_2$ dissociation on various other transition metal 
surfaces \cite{Ren89,Res94,Ber92,But94,Aln89,But95,Dix94}. 
Such a behaviour is usually found for atomic or non-dissociative molecular 
adsorption in which the particles have to dissipate their excess energy
to the substrate in order to remain on the surface \cite{Zan88}. Since
the energy transfer to the substrate becomes less efficient at higher 
energies, the sticking probability for atomic or molecular adsorption
decreases with increasing energy. Consequently a similiar process has
been invoked to explain a decreasing sticking coefficient in {\em dissociative}
adsorption, namely the precursor mechanism:
before dissociation the molecules are temporarily trapped
in a molecular adsorption state, the so-called precursor state.
This trapping probability which decreases with increasing energy 
is supposed to be the rate-determining step.

\begin{figure}[tb]
\unitlength1cm
\begin{center}
   \begin{picture}(10,11.0)
\put(-1.5,-.5){ {\epsfysize=12.cm  
          \epsffile{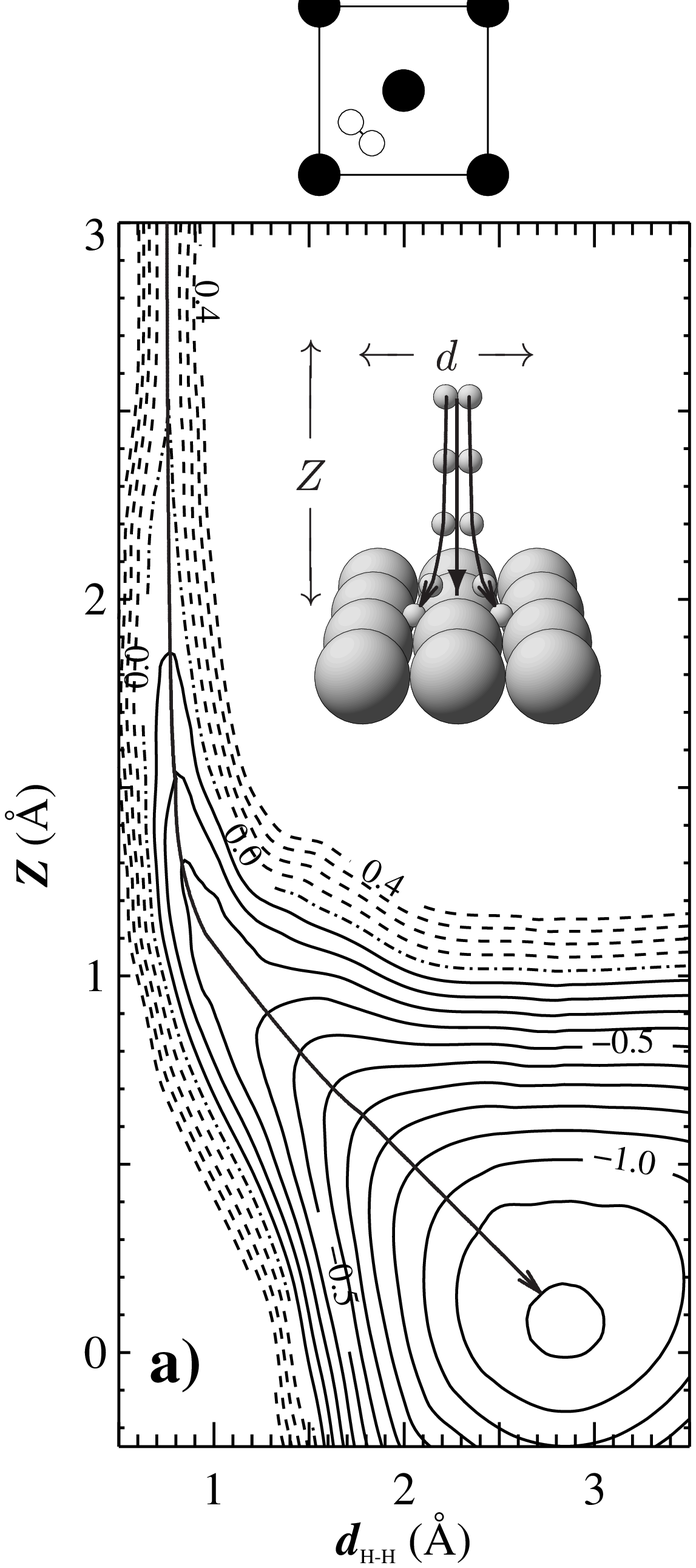}} }
\put(2.5,-.5){ {\epsfysize=12.cm  
          \epsffile{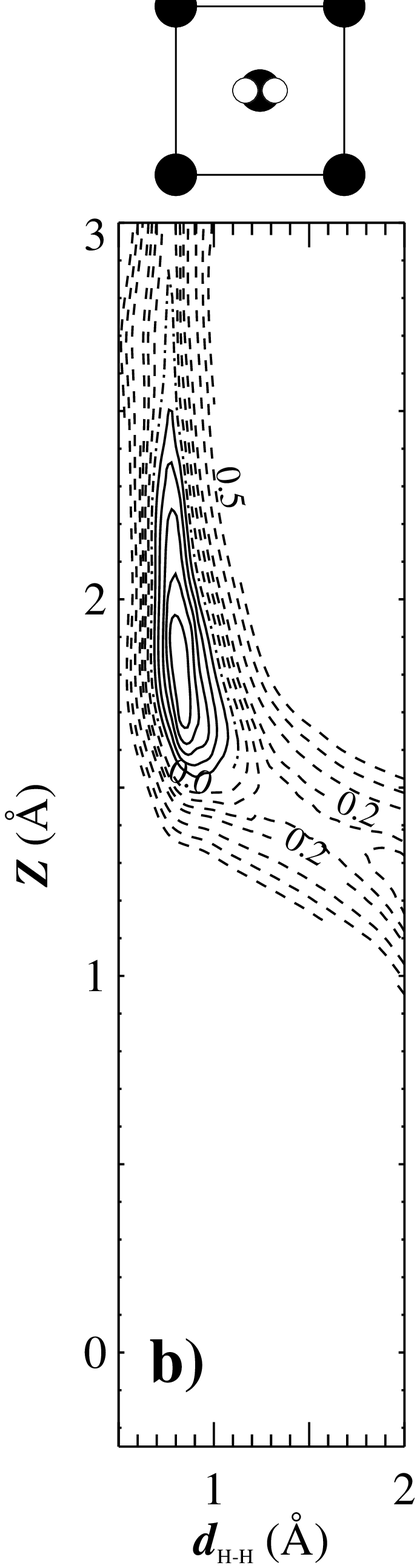}} }

   \end{picture}

\end{center}
   \caption{Contour plots of the PES along two two-dimensional cuts 
            through the six-dimensional coordinate space of 
            H$_2$/Pd\,(100), so-called elbow plots, determined by GGA 
            calculations \protect\cite{Wil95,Wil96PRB}. The coordinates 
            in the figure are the H$_2$ center-of-mass distance 
            from the surface $Z$ and the H-H interatomic distance $d$.
            The dissociation process in the $Zd$ plane is illustrated
            in the inset. The lateral H$_2$ center-of-mass coordinates 
            in the surface unit cell and the orientation of the molecular 
            axis, i.e. the coordinates $X$, $Y$, $\theta$, and $\phi$  
            are kept fixed for each 2D cut and depicted above the 
            elbow plots. Energies are in eV per H$_2$ molecule.
            The contour spacing in a) is 0.1~eV, while it is 0.05~eV in b).}

\label{h2pdelbow}
\end{figure}

However, for hydrogen adsorption the large mass mismatch between adsorbate
and substrate should make the energy transfer process inefficient.  
For the system H$_2$/W(100)--c(2$\times$2)Cu, e.g., it was shown \cite{But95} 
that for a hydrogen molecule impinging on a metal substrate 
the energy transfer to substrate phonons is much too small to account 
for the high sticking probability at low kinetic energies.

Furthermore, density-functional theory calculations within the GGA
found that there exist non-activated as well as activated paths to 
dissociative adsorption in the system H$_2$/Pd(100), but the calculations
gave no indication of any molecular adsorption well, i.e. any precursor 
state \cite{Wil95,Wil96PRB}. 
Two elbow plots of the H$_2$/Pd(100) PES determined in these calculations 
are shown in Fig.~\ref{h2pdelbow}. Fig.~\ref{h2pdelbow}a) demonstrates
that the dissociative adsorption of H$_2$ on Pd(100) is non-activated,
i.e. reaction pathways without any hindering barrier exist. For the
conditions  of Fig.~\ref{h2pdelbow}b), where the molecule approaches the 
surface at the on-top site, a barrier of approximately 0.15~eV
exists. There seems to be a molecular adsorption well in front of the
on-top site, but detailed calculations have shown that this well
does not correspond to a local minimum of the PES \cite{Wil96PRB}.
It is rather a saddle point of the PES in the multi-dimensional configuration
space because the molecule can still follow a purely attractive
path to dissociative adsorption if its center-of-mass degrees of freedom
are allowed to relax.

According to this theoretical results,  
the precursor mechanism can not be operative in the H$_2$/Pd(100)
system. As an alternative explanation it had been suggested that a decreasing
sticking coefficient could be caused by a steering effect in a direct 
non-activated adsorption process by King twenty years ago \cite{King78}, 
but there had been no theoretical confirmation whether this mechanism could 
be efficient enough. In classical stochastic molecular dynamics simulations
it was demonstrated how the anisotropy and corrugation could lead to a
focusing of the dissociating molecules to reactive sites, but the energy
dependence of this process was not analysed \cite{Kara90}. 
Furthermore, in two-dimensional dynamical treatments of the
H$_2$/Pd(100) system no steering effect was observed \cite{Sch92,Dar92b,Bre94}.

\begin{figure}[tb]
\unitlength1cm
\begin{center}
   \begin{picture}(10,6.5)
{   \rotate[r]{\epsfysize=8.cm  
          \epsffile{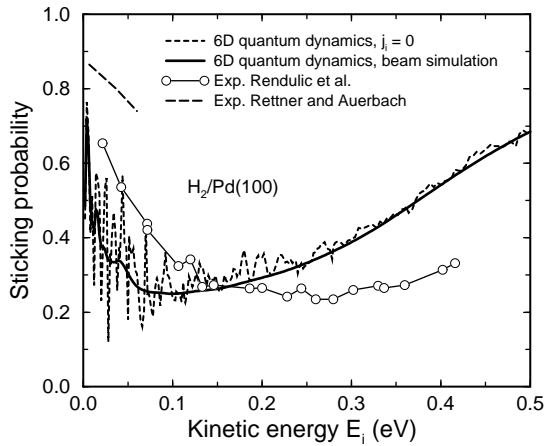}} }

   \end{picture}
\end{center}
   \caption{Sticking probability versus kinetic energy for
a hydrogen beam under normal incidence on a Pd(100) surface.
Theory: six-dimensional quantum calculations for H$_2$ molecules initially 
in the rotational and vibrational ground state (dashed line)
and with an initial rotational and energy distribution 
adequate for molecular beam experiments (solid line) \protect\cite{Gro95PRL}.
H$_2$ molecular beam adsorption experiment under normal incidence
(Rendulic {\it et al.}~\protect\cite{Ren89}): circles;
H$_2$ effusive beam scattering experiment with an incident angle of
of $\theta_i = 15^{\circ}$
(Rettner and Auerbach~\protect\cite{Ret96}): long-dashed line. }
\label{h2pdstick}
\end{figure}

The first dynamical evidence that the co-existence of non-activated with
activated pathways to dissociative adsorption can lead to a initially
decreasing sticking probability was found in three-dimensional quantum 
dynamical calculations \cite{Gro95JCP}. In this study a model PES with 
features derived from the GGA calculations of H$_2$/Pd(100) \cite{Wil95}
was used. However, still large quantitative discrepancies to the 
experiment remained, as far as the 
relevant energy range in which steering is operative was concerned.

Using a parametrization of the {\it ab initio} PES \cite{Wil95,Wil96PRB} 
based on 250 different configurations,
six-dimensional dynamical calculations of the 
dissociative adsorption and associative desorption of H$_2$/Pd(100) 
have been performed \cite{Gro95PRL}. These calculations were indeed the first 
in which all six degrees of freedom of the hydrogen molecule were treated 
quantum dynamically. Figure~\ref{h2pdstick} presents the 6D results
of the sticking probability of H$_2$/Pd(100) under normal incidence. 
These results are compared to the H$_2$ molecular
beam adsorption experiment by Rendulic, Anger and Winkler \cite{Ren89}.
Furthermore, the experimental results of Rettner and Auerbach for a
reflection beam experiment under an angle of incidence of 
$\theta_i = 15^{\circ}$ \cite{Ret96} are plotted.

First of all, a strong oscillatory structure is evident in
the quantum results. Such oscillations have also been found in
4D wave-packet calculations of H$_2$/W(100) \cite{Kay95} which is also a
reactive system; those were, however, smaller.
It has been shown that the energetic location of most of the peaks,
in particular at low kinetic energies, can be related to the opening
of new scattering channels with increasing kinetic energies, i.e. they
correspond to threshold effects \cite{Gro96PRL,Gro96CPLb}. In the case
of the diffraction channels, the energetic position of the peaks is thus
fully determined by the geometry of the surface. Their size depends
sensitively on the symmetry of the initial conditions. For normal
incidence they are much more pronounced than for non-normal incidence
because the number of degenerate, symmetrically equivalent scattering
channels is reduced for non-normal incidence \cite{Gro98PRB,Gro96CPLb}.
Holloway and coworkers have recently shown \cite{Hol96} that the size
of the oscillations also depends on the specific topology of the PES,
in particular on the existence of possible resonance states in front
of the surface.

\begin{figure*}[t]
\unitlength1cm
\begin{center}
   \begin{picture}(20,8.0)
\put(-1.5,0.){\rotate[l]{\epsfxsize=7.5cm  
          \epsffile{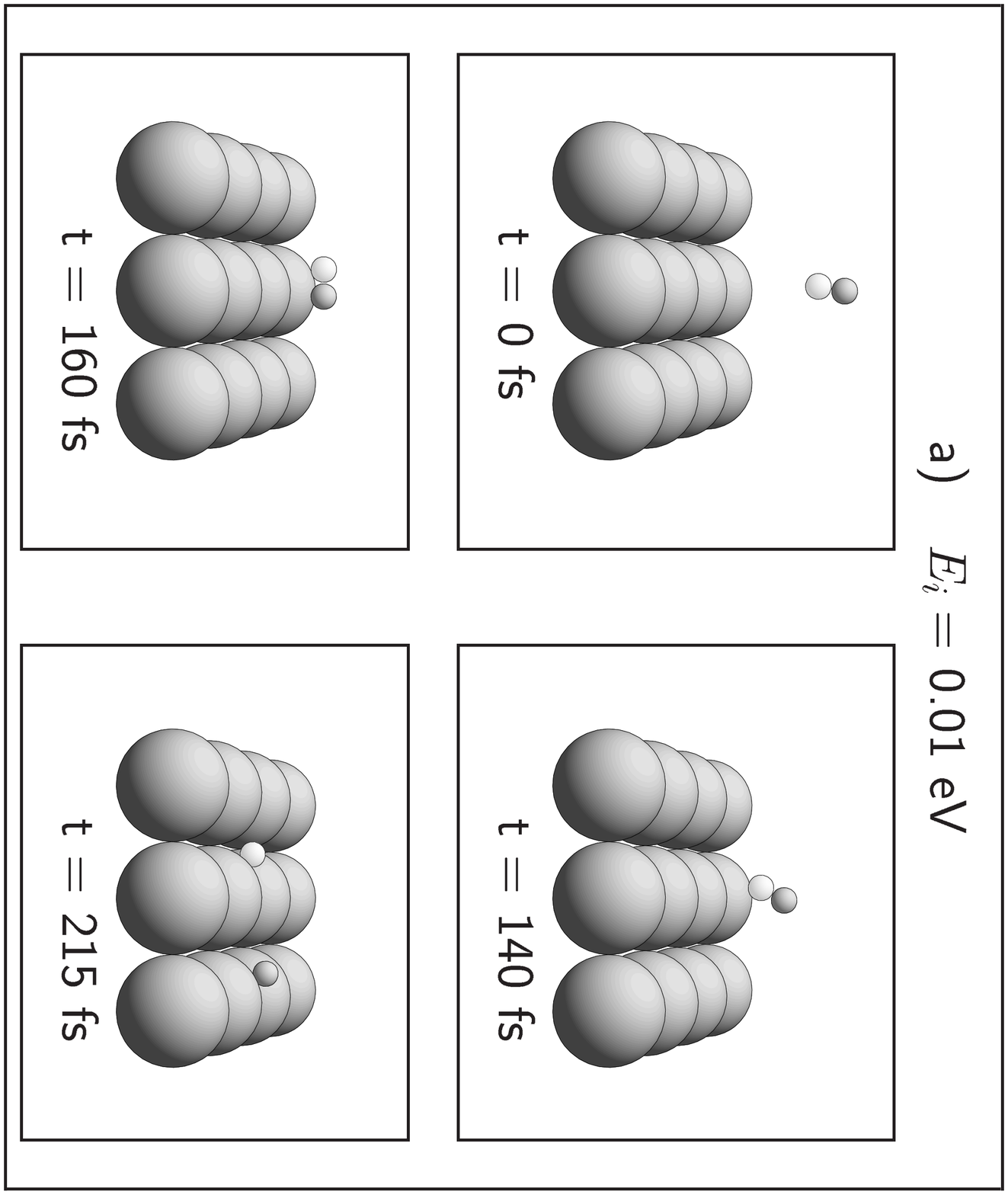}} }
\put(8.0,0.){\rotate[l]{\epsfxsize=7.5cm  
          \epsffile{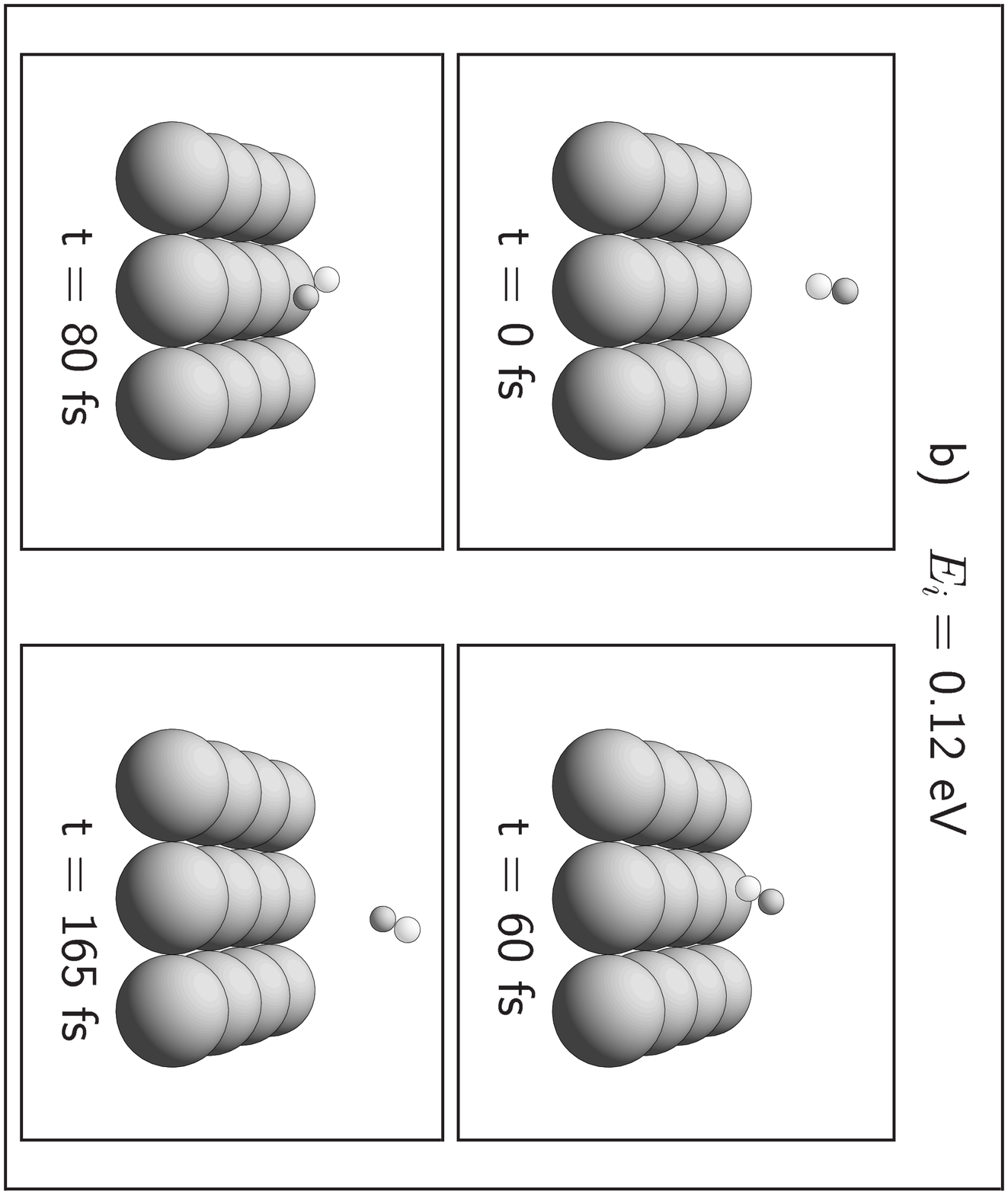}} }
   \end{picture}

\end{center}
   \caption{Snapshots of {\it ab initio} molecular dynamics trajectories
            for H$_2$ molecules impinging on Pd(100) 
            in order to illustrate the steering effect. The molecules
            are initially not vibrating and not rotating. For both
            trajectories the initial conditions are the same, except 
            for the  kinetic energy which is 0.01~eV in a) and 
            0.12~eV in b). 
                }

\label{h2pdtraj}
\end{figure*}

Rettner and Auerbach have  searched for these oscillations in a 
reflection experiment using a nearly effusive beam \cite{Ret96,Ret96PRL}.
This kind of experiment allows a high energy resolution of the total 
scattering probability \cite{Ret93CP}. They have not found any 
indication of these oscillations. However, this is not an easy experimental
task since surface impurities and the increasing hydrogen coverage
during the scattering experiment destroy the coherence of the scattering
event and suppress the oscillations. Furthermore, they have used an
angle of incidence of 15$^{\circ}$ instead of normal incidence which also
causes a suppression of the oscillations \cite{Gro98PRB}, since 
the symmetry of the initial conditions has a decisive influence
on the magnitude of the oscillations. Simulated
time-of-flight (TOF) distributions for H$_2$ scattered at Pd(100) 
with the same  angle of incidence as used in the experiment are very
close to the experimental TOF distributions \cite{Gro98PRB}. The results
of Rendulic, Anger and Winkler \cite{Ren89} do not show any strong 
oscillatory structure either, but for an additional reason.
An experimental molecular beam does not correspond
to a monoenergetic beam in one specific quantum state. 
If one assumes an energy spread and a distribution of internal
molecular states typical for a beam experiment, the oscillations
are almost entirely smoothed out in the 6D quantum results
(solid line in fig.~\ref{h2pdstick}).

The general agreement between theory and the experiments is satisfactory.
As noted above, the {\it ab initio} PES does not have any molecular
adsorption well, i.e., no precursor state. In addition, due to the fact
that the substrate is kept fixed, no energy transfer to the surface is 
taken into account in the calculations. Still the initial decrease of the 
sticking probability found in the experiments is well-produced in the 
quantum dynamical calculations. 
Such an initial decrease of the sticking probability has also been found in 
classical molecular dynamics calculations on the same PES 
\cite{Gro98PRB,Gro97Vac} and in quantum and classical calculations of the 
sticking probability of H$_2$/W(100), too \cite{Kay95}.
Hence the underlying microscopic mechanism that is responsible for the 
decrease can also be investigated by analysing classical trajectories.
This analysis yields that the large sticking probability at low kinetic
energies is caused by the steering effect that becomes less efficient
at higher kinetic energies which then leads to the decrease of the
dissociation probability.

Figure~\ref{h2pdtraj} illustrates this general mechanism
by showing snapshots of two {\it ab initio} molecular 
dynamics runs. For both trajectories initially non-vibrating and
non-rotating molecules impinge on the surface under the same conditions  
except for the initial kinetic energy which is 0.01~eV in 
Fig.~\ref{h2pdtraj}a) and 0.12~eV in Fig.~\ref{h2pdtraj}b). Far 
away from the surface the molecular axis is oriented almost perpendicular
to the surface. In such an orientation the molecule cannot dissociate, the
interaction with the surface is repulsive. The PES is, however, strongly
anisotropic, and there are forces acting on the molecule to orient
it to a parallel configuration to the surface. If the molecule is slow enough,
as in Fig.~\ref{h2pdtraj}a), it can indeed complete the rotation into this 
favorable orientation before hitting the repulsive wall of the potential.
Close to the surface, after 160~fs, the molecule has turned parallel to
the surface and directly dissociates from this configuration. This process
becomes less efficient at higher kinetic energies which is demonstrated
in Fig.~\ref{h2pdtraj}b) where the initial kinetic energy is twelve times
larger. Of course the same forces act upon the molecule, but now the 
molecule is too fast to be fully re-oriented. It hits the surface
in an unfavorable configuration in which the interaction is repulsive.
At the classical turning point there is a very rapid rotation corresponding
to a flip-flop motion, and then the molecule is scattered back into 
the gas-phase rotationally excited.
By further increasing the kinetic energy, the molecules will eventually
have enough energy to directly traverse the barrier region which causes
the final increase in the sticking probability (see Fig.~\ref{h2pdstick}).

The events that are depicted in Fig.~\ref{h2pdtraj}, direct dissociation
or direct reflection, only correspond to the two limits of the possible
outcome of a scattering event. A slow molecule approaching the surface
can start rotating caused by the anisotropy of the PES
without directly dissociating. Due to the conversion
of translational energy into rotational energy it might lose
so much translational energy that it cannot escape into
the gas phase again. The corrugation of the surface can also cause
motion parallel to the surface. In such a state the molecule is trapped
into a dynamical precursor, which is {\em not} caused by energy transfer
to the substrate but by energy transfer to rotational and kinetic 
energy of the molecule parallel to the surface. In {\it ab initio}
molecular dynamics calculations adsorption events have been found
in which the molecule spent more than 5~ps in front of the surface 
before dissociating \cite{Gro97Vac}. Note that such events correspond to 
the existence of metastable states in quantum scattering. These are
very sensitive to the specific topology of the PES \cite{Hol96}.

\begin{figure}[tb]
\unitlength1cm
\begin{center}
   \begin{picture}(10,6.0)
      \includegraphics{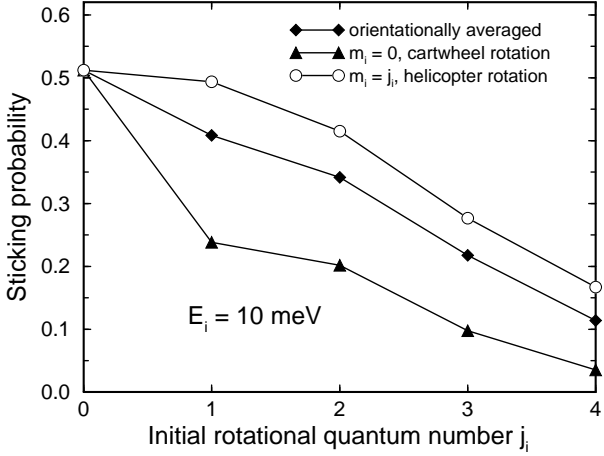}
   \end{picture}

\end{center}
   \caption{Sticking probability versus initial rotational quantum state $j_i$
   for the system H$_2$/Pd(100).
   Diamonds: orientationally averaged sticking probability 
   (eq.~\protect\ref{ave}),
   triangles: $m_i = 0$ (cartwheel rotation), 
   circles: $m_i = j_i$ (helicopter rotation). 
   The initial kinetic energy is $E_{i} = 10$~meV.}

\label{steric}
\end{figure}

Both the trapping into a molecular adsorption state by energy dissipation
to the substrate and the suppression of the steering effect
lead to a sticking probability that decreases with increasing kinetic
energy. It is therefore desirable to find experimental properties
that allow an unambiguous distinction between these two different
mechanisms. As such a property the dependence of the sticking probability
on the initial rotational motion of the molecules has been suggested
\cite{Gro96SSb}. One of the properties of the PES that leads to
steering, the anisotropy of the potential, also causes a suppression
of the sticking probability when the molecules are rapidly rotating.
Molecules with a high angular momentum will rotate 
out of a favorable orientation towards dissociative adsorption during the 
time it takes to break the molecular bond. This is shown in
Fig.~\ref{steric} where the 6D sticking probability for a kinetic
energy of $E_i = 10$~meV as a function of the initial rotational state 
is plotted. The diamonds correspond to the orientationally averaged 
sticking probability 
\begin{equation}\label{ave}
\bar S_{j_i} (E) \ = \ \frac{1}{2j_i +1} \ \sum_{m_i = -j_i}^{j_i} \ 
S_{j_i,m_i} (E). 
\end{equation}
The strong decrease with increasing initial rotational quantum number $j_i$
is caused by a suppression of the steering effect. 
Fast rotating molecules cannot be focused into a favorable orientation 
towards adsorption, they will be reflected from the surface.

\begin{figure}[tb]
\unitlength1cm
\begin{center}
   \begin{picture}(10,6.5)
\centerline{   \rotate[r]{\epsfysize=8.cm  
          \epsffile{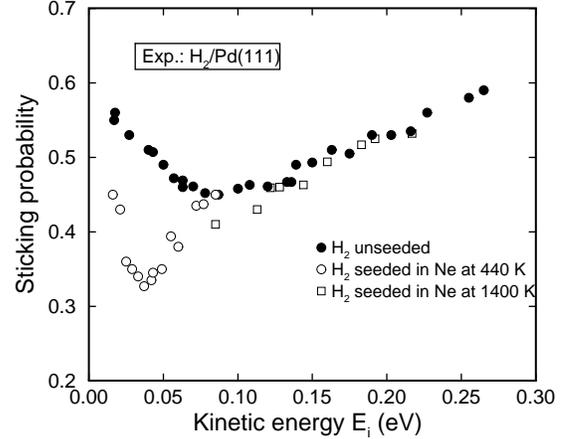}} }
   \end{picture}

\end{center}
   \caption{Experimental sticking probability versus kinetic energy for
a H$_2$ beam under normal incidence on a Pd(111) surface.
The seeded beams have larger rotational energies at identical kinetic
energies than the unseeded beam (after \protect\cite{Beu95}).
}

\label{h2pd111exp}
\end{figure}

It is not easy experimentally to prepare a molecular beam in
a single quantum state. Still, this rotational hindering of the steering 
effect has actually been confirmed for 
H$_2$/Pd(111) \cite{Beu95,Gos97} and also for H$_2$/Pt(110) \cite{Beu96}. 
By seeding techniques the translational 
energy of a H$_2$ beam can be changed in a nozzle experiment without 
altering the rotational population of the beam \cite{Beu95,Gos97}.
Also the different sticking probabilities of para- and $n$-hydrogen
beams have been used to extract information about rotationally
resolved results \cite{Beu96}.
Figure~\ref{h2pd111exp} shows the results of a seeded beam
experiment for H$_2$/Pd(111) by Beutl and coworkers \cite{Beu95}. 
The seeded beams have larger rotational 
energies than the unseeded beam at the same kinetic energies. It is
said, they are rotationally hot. And indeed, especially at low kinetic
energies the predicted strong suppression of the sticking probability
by the rotational hindering is found, see Fig.~\ref{h2pd111exp}.

According to the principle of detailed balance, the suppresion of the 
sticking probability by the rotational hindering should be reflected
by a population of rotational excited states in desorption 
which is lower than expected for molecules in thermal equilibrium 
with the surface temperature.
This so-called rotational cooling has indeed been found for H$_2$ 
molecules desorbing from Pd(100) \cite{Sch91a,Sch91b} and is also well 
reproduced by the six-dimensional quantum dynamical 
calculations \cite{Gro95PRL}.

In the precursor mechanism the molecules are assumed to be first
trapped in a physisorption state. Such a state is caused by 
van-der-Waals forces which depend only weakly on the orientation
of the molecule. Hence there are only small directional forces,
and molecules adsorbed in such a physisorption state are able to rotate 
almost freely \cite{Zan88,And82}. The trapping 
probability into the physisorption 
state and thus the sticking probability in the precursor model should 
therefore be almost independent of the initial rotational state, in contrast 
to the steering mechanism. In another  series
of experiments Beutl and coworkers have measured the dependence of
the sticking probabilities on the initial rotational energies
in the ``classical'' precursor systems N$_2$/W(100), CO/FeSi(100) 
and O$_2$/Ni(111) \cite{Beu97}. They found no discernible 
influence of the rotational motion on the sticking coefficient in these 
systems. Hence the rotational dependence might indeed serve as a
means to distinguish between the precursor mechanism and the
steering effect.

So far all the results presented were averaged over the azimuthal
quantum number $m$, i.e, the results corresponded to orientationally
averaged properties. Now the H$_2$/Pd(100) PES is strongly anisotropic
with regard to the molecular orientation. The most favourable
configuration towards dissociative adsorption is with the molecular
axis parallel to the surface. Molecules that hit the surface in an
upright position cannot dissociate, they are reflected back into the 
gas-phase. It is true that quantum mechanics does not allow for
non-rotating, oriented molecules in the gas-phase, however, rotating
molecules can show a preferential orientation. 
Fig.~\ref{steric} also shows the effect 
of molecular orientation on the sticking probability. Molecules with azimuthal
quantum number~$m = j$ have their axis preferentially oriented parallel to the 
surface. These molecules rotating in the so-called helicopter fashion  
dissociate more easily than molecules rotating in the cartwheel fashion 
($m = 0$) with their rotational axis preferentially parallel to the
surface  since the latter have a high probability hitting the
surface in an upright orientation in which they cannot dissociate.

Experimentally it is hard to align a molecular beam of hydrogen.
Again one can study the time-reverse process, the associative desorption.
By laser-induced fluorescence (LIF) it is possible to measure
the rotational alignment parameter $A_0^{(2)}(j)$ \cite{Gre83}, 
which is given by
\begin{equation}
A_0^{(2)} (j) \ = \ \left\langle \frac{3J_z^2 \ - \ {\bf J}^2}{{\bf J}^2}
\right\rangle_j 
\end{equation}
$A_0^{(2)}(j)$ corresponds to the quadrupole moment of the orientational
distribution and assumes values of $-1 \ \leq \ A_0^{(2)} (j) \ \leq 2$.
Molecules rotating preferentially in the cartwheel fashion have an
alignment parameter $A_0^{(2)} (j) < 0$, for molecules rotating 
preferentially in the helicopter fashion $A_0^{(2)} (j) > 0$.

\begin{figure}[tb]
\unitlength1cm
\begin{center}
   \begin{picture}(10.,6.0)
      \includegraphics{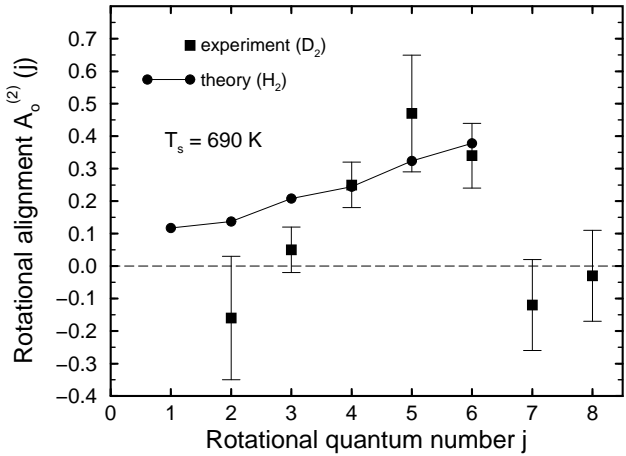}
   \end{picture}

\end{center}
   \caption{Rotationally alignment of hydrogen molecules desorbing
       from a Pd(100) surface. 
       Boxes: experimental results for D$_2$ \protect\cite{Wet96}. 
       Circles: 6-D quantum dynamical calculations for H$_2$
       (from \protect\cite{Gro96Prog}).  
       }

\label{rotaligndes}
\end{figure}

Experiments have in fact found a preferential population of the
helicopter mode in the desorption of D$_2$/Pd(100) \cite{Wet96}.
Figure~\ref{rotaligndes} shows a comparison of these experimental
results for D$_2$ with the six-dimensional calculation of the 
rotational alignment of H$_2$ desorbing from a
Pd(100) surface at a surface temperature of $T_s = 690$~K \cite{Gro96Prog}. 
Experimentally there is no big difference in the alignment factors 
as a function of the rotational quantum number $j$ between H$_2$ and 
D$_2$ \cite{Zach97}, hence the comparison is meaningful. 
The agreement for $j \le 6$ is quite satisfactory. Indeed the molecules
desorb preferentially with their molecular axis parallel to the surface
thus reflecting the anisotropy of the PES. Due to computational
restrictions the rotational alignment could only be calculated for
$j \le 6$. For $j = 7$ and $j =8$ the experiments show a vanishing
alignment within the error bars which is rather surprising. It was
argued that for these high rotational states an energy transfer from
the rotations to the reaction coordinate could suppress the effect
of the potential anisotropy \cite{Wet96}. Certainly these results 
deserve further clarification.

The dependence of adsorption and desorption 
on kinetic energy, molecular rotation and orientation
\cite{Gro95PRL,Gro96SSb}, molecular vibration \cite{Gro96CPLa}, ro-vibrational
coupling \cite{Gro96Prog}, angle of incidence \cite{Gro98PRB}, 
and the rotationally elastic and inelastic 
diffraction of H$_2$/Pd(100) \cite{Gro96CPLb} have been studied so far
by six-dimensional {\it ab initio} dynamics calculations
on the same PES. The results of these calculations have been
compared to a number of independent experiments, and they are
at least in semi-quantitative agreement with all of these experiments.
Recently six-dimensional {\it ab initio} quantum dynamical calculations
of H$_2$/Pd(100) have been performed \cite{Eich98Diss,Eich98pre}
which were based on a much larger
set of DFT-GGA calculations \cite{Eich98}. This allowed a more detailed
analytical representation of the PES and significantly improved the
agreement between theoretical and experimental sticking probabilitities
\cite{Eich98Diss,Eich98pre}. 
This shows that that {\it ab initio} dynamics calculations are
capable of adequately describing the hydrogen dissociation
on transition metal surfaces. In addition, the difference 
between six-dimensional {\it ab initio} quantum dynamics 
and classical dynamics has been
addressed \cite{Gro97Vac}. As in the case of the 3D-calculations
of the hydrogen dissociation on Cu(111) (see Fig.~\ref{qmclass}) the crucial
difference between classical and quantum dynamics is not due to
tunneling, but to zero-point effects. However, caused by the higher 
dimensionality of the 6D-calculations the number of zero-point vibrational
modes perpendicular to the reaction path is larger which makes the difference
between classical and quantum results much more pronounced. 
This shows that a proper inclusion of zero-point effects is required
in the treatment of hydrogen dynamics. I will address this issue in
more detail in the next section about the hydrogen dissociation on
an adsorbate covered surface.

It is also well-known that palladium is able to absorb large amounts
of hydrogen in the bulk and can therefore be used for hydrogen storage
\cite{Chr88}. The process of how hydrogen might enter the bulk has been
the issue of {\it ab initio} quantum dynamics studies employing two- and
three-dimensional wave-packet calculations by Olsen 
{\it et al.} \cite{Ols97a,Ols97b}. In order to determine the effect of
surface motion on direct subsurface absorption of H$_2$/Pd(111)  
they have calculated a GGA-PES depending on two hydrogen and one
palladium degree of freedom \cite{Ols97b}. They found that indeed 
the surface degrees of freedom play an important role at low kinetic 
energies as surface relaxation can lower the effective barrier towards 
direct subsurface absorption substantially. Still this barrier 
is too large to explain the experimentally observed direct subsurface
absorption \cite{Gdo87}, probably due to the restricted dimensionality 
of these calculations, so that higher-dimensional calculations are
desirable.


\subsection{Dissociation on an adsorbate covered metal surface}

The presence of an adsorbate on a surface can profoundly change the 
surface reactivity. An understanding of the underlying mechanisms
and their consequences on the reaction rates is of decisive importance
for, e.g., designing better catalysts. 
Traditionally an ``trial and error'' approach
was used to improve the activity of a catalyst by adding some
substances. Only recently this problem has been addressed theoretically
by {\it ab initio} methods \cite{Bes98}. 

On Pd(100) it is experimentally well-known that the presence of sulfur 
leads to a large reduction of the hydrogen dissociation 
probability \cite{Ren89,Bur90}, it ``poisons'' the dissociation. 
While at the clean surface the dissociation
probability is about 60\% for a kinetic energy of $E_{\rm i} = 0.05$~eV,
at the sulfur-covered surface it drops below 1\% at the same
energy \cite{Ren89}.

\begin{figure}[tb]
\unitlength1cm
\begin{center}
   \begin{picture}(10,8.0)
\centerline{   {\epsfxsize=8.cm  
          \epsffile{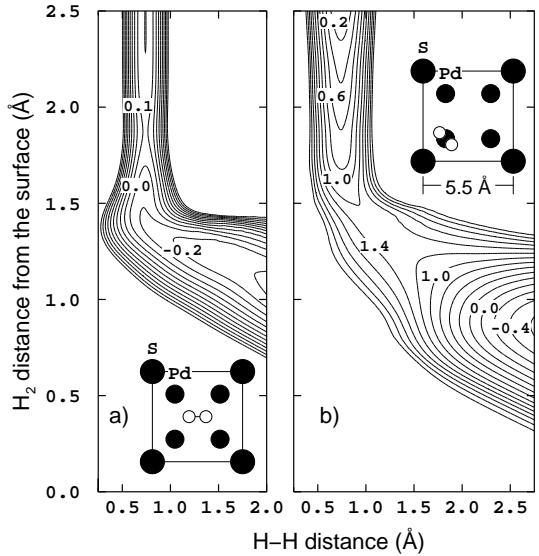}} }
   \end{picture}
\end{center}
   \caption{Contour plots of the PES along two two-dimensional cuts 
            through the six-dimensional coordinate space of 
            H$_2$/S(2$\times$2)/Pd\,(100). The insets show the
            orientation of the molecular axis and the lateral
            H$_2$ center-of-mass coordinates, i.e. the coordinates
            $X$, $Y$, $\theta$, and $\phi$. The coordinates 
            in the figure are the H$_2$ center-of-mass distance 
            from the surface $Z$ and the H-H interatomic distance $r$. 
            Energies are in eV per H$_2$ molecule.
            The contour spacing in Fig.~\protect\ref{elbow}a  is 0.1~eV,
            while in Fig.~\protect\ref{elbow}b it is 0.2~eV.
            Fig.~\protect\ref{elbow}a corresponds to the minimum energy
            pathway.}
\label{elbow}
\end{figure}

DFT-GGA calculations have shown that hydrogen dissociation on 
sulfur-covered Pd(100) is still exothermic,
however, the dissociation is hindered by the formation of energy barriers
in the entrance channel of the potential energy surface 
(PES) \cite{Wil96S}.
Figure~\ref{elbow} shows two elbow plots of an analytical fit
to the GGA-PES of H$_2$ at the (2$\times$2) sulfur
covered Pd(100) surface for fixed
molecular orientation and lateral center-of-mass coordinates of the
molecule. 
The minimum energy path is shown in Fig.~\ref{elbow}a. The molecule
approaches the surface above the fourfold hollow site. This is the
site where the H$_2$ molecule is furthest away from the sulfur atoms 
on the surface. An analysis of the electronic structure calculations
has revealed that the building up of the barrier at this position is
not related to a direct interaction between sulfur and the hydrogen 
molecule. Instead, it is caused by the sulfur-mediated downshift of
the Pd $d$~bands at the surface \cite{Wil96S,Wei97} which
leads to a population of anti-bonding molecule-surface 
states \cite{Ham95PRL}.

Closer to the sulfur atoms the PES becomes strongly repulsive.
This is illustrated in Fig.~\ref{elbow}b. While the dissociation 
path over the Pd on-top position on the clean surface is hindered by a 
barrier of height 0.15~eV \cite{Wil96PRB}, the adsorbed sulfur leads 
to an increase in this barrier height to 1.3~eV. Directly above the 
sulfur atoms the barrier towards dissociation even increases to values 
larger than 2.5~eV~\cite{Wei97} for molecules oriented parallel to
the surface. Here direct interaction between the sulfur atom and 
the H$_2$ molecule is responsible for the repulsion and the high 
barrier \cite{Wei97}.

\begin{figure}[tb]
\unitlength1cm
\begin{center}
   \begin{picture}(10,6.5)
\centerline{   \rotate[r]{\epsfysize=8.cm  
          \epsffile{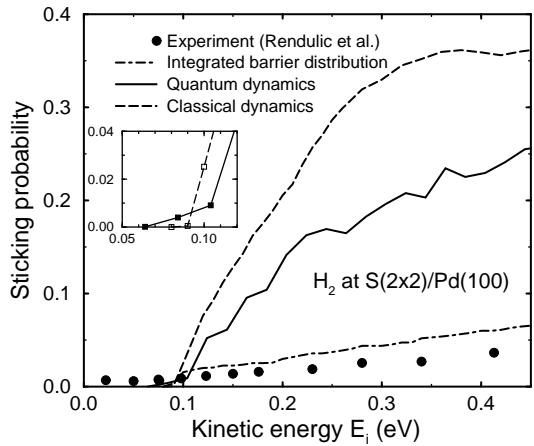}} }
   \end{picture}

\end{center}
   \caption{Sticking probability versus kinetic energy for
a H$_2$ beam under normal incidence on a S(2$\times$2)/Pd(100) surface.
Full dots: experiment (from ref.~\protect\cite{Ren89});
Dashed-dotted line: Integrated barrier distribution,
which corresponds to the sticking probability in the hole 
model \protect\cite{Kar87};
Solid line: Quantum mechanical results for molecules initially in the
rotational and vibrational ground-state;
Dashed line: Classical results for initially non-rotating and non-vibrating
molecules. The inset shows the quantum and classical results at low 
energies.}

\label{stick}
\end{figure}

On the analytical representation of the PES of H$_2$ at the
(2$\times$2) sulfur covered surface six-dimensional dynamical
calculations have recently been performed \cite{Gro98PRL}.
To my knowledge, this represents the most corrugated system of 
dissociative adsorption studied dynamically so far. Figure \ref{stick} 
compares the quantum and classical results for the sticking 
probability as a function of the kinetic energy of the 
incident H$_2$ beam with the experiment \cite{Ren89}.
In addition, also the integrated barrier distribution $P_b(E)$,
\begin{eqnarray} \label{barr}
P_b (E) & = & \frac{1}{2\pi A} \ \int 
\Theta (E - E_{\rm b} (\theta, \phi, X, Y)) \nonumber \\
& & \times \ \cos \theta d\theta \ d\phi \ dX \ dY
\end{eqnarray}
is plotted. Here $\theta$ and $\phi$ are the polar and azimuthal
orientation of the molecule, $X$ and $Y$ are the lateral coordinates of 
the hydrogen center-of-mass. $A$ is the area of the surface unit cell.
Each quadruple defines a cut through the six-dimensional space (see
Fig.~\ref{elbow} for examples), and $E_{\rm b}$ is the minimum energy barrier 
along such a cut. The function $\Theta$ is the Heavyside step function. 
The quantity $P_b(E)$ is the fraction 
of the configuration space for which the barrier towards dissociation
is less than $E$; it corresponds to the sticking probability
in the classical sudden approximation or the so-called 
``hole model'' \cite{Kar87}.

The calculated sticking probabilities are significantly larger than the
experimental results. The onset of dissociative adsorption at 
$E_{\rm i} \approx 0.12$~eV, however, is well reproduced by the calculations. 
This onset is indeed also in agreement
with the experimentally measured mean kinetic energy of hydrogen 
molecules desorbing from sulfur covered Pd(100) \cite{Com80}.
The theoretical results are relatively insensitive to variations in
the barrier heights of the order of 0.2~eV \cite{Gro98PRL}. It might
therefore well be that uncertainties in the experimental determination
are responsible for the difference. The exact sulfur coverage in the 
experiment was not very well characterized. The sulfur adlayer was obtained 
by simply heating up the sample which leads to segregation of bulk sulfur 
at the surface. The sulfur coverage was monitored
through the ratio of the Auger peaks S$_{132}$/Pd$_{330}$ \cite{Ren89}.
The set of experimental data shown in Fig.~\ref{stick} were obtained for 
half of the maximum segregation coverage of sulfur on Pd(100).
This saturation coverage is assumed to be roughly $\Theta_S \approx 0.5$.
Apart from uncertainties in this determination of the sulfur coverage
also subsurface sulfur might have influenced the measurements.
In ref.~\cite{Bur90} a linear decrease of the hydrogen saturation 
coverage with increasing sulfur coverages was found. At a sulfur coverage
of $\Theta_{\rm S} = 0.28$ hydrogen adsorption should be completely suppressed.
This infers a site-blocking effect which is at variance
with the {\it ab initio} calculations of the PES for this 
system \cite{Wil96S,Wei97}. However, these seemingly contradicting
results and also the discrepancy between calculated and measured
sticking probabilities could be reconciled if subsurface sulfur (which is not
considered in the calculations but which might well be present in the
experimental sample) has a decisive influence on the hydrogen adsorption
energies. It is certainly desirable that the effect of subsurface
sulfur on the hydrogen adsorption in this system will be investigated.

\begin{table}
\begin{center}
\begin{tabular}{|c|l|}
\hline
mode & ZPE (eV)\\
\hline
H-H vibration & 0.253 \\
polar rotation & 0.016 \\
azimuthal rotation & 0.013 \\
translation perpendicular to molecular axis & 0.027 \\
translation parallel to molecular axis & 0.027 \\
\hline
sum & 0.336 \\
\hline
\end{tabular}
\caption{Zero-point energies (ZPE) of the H$_2$ molecule at the
minimum barrier position. The H$_2$ configuration
corresponds to the situation of Fig.~\protect\ref{elbow}a. 
The gas-phase zero-point energy of H$_2$ is
$\frac{1}{2} \hbar \omega_{\rm gas} =$~0.258~eV.
}
\label{zeros}
\end{center}

\end{table}

Figure \ref{stick} shows furthermore that the classical molecular
dynamics calculations over-estimate the sticking probability 
of H$_2$ at S(2$\times$2)/Pd(100) compared to the quantum results. 
There are two important quantum phenomena not taken into account 
by classical calculations: tunnelling and zero-point effects.
For energies smaller than the minimum barrier height $E_{\rm b}$ the classical
sticking probability is of course zero, whereas the quantum results
still show some dissociation due to tunneling, as the inset of 
Fig.~\ref{stick} reveals. But for energies larger than
$E_{\rm b}$ the classical sticking probability rises to 
values which are almost 50\% larger than the quantum sticking
probabilities. Tunnelling increases the quantum transmission probability 
with respect to the classical result, hence tunnelling would have the 
opposite effect as observed in Fig.~\ref{stick}. Thus only zero-point 
effects can be responsible for the large difference \cite{Gro98PRB,Gro97Vac}.  

Since the hydrogen molecular bond is still almost intact at the 
minimum barrier position, the H-H vibrational frequency should not
be too different from its gas-phase value there. However, the
minimum barrier is localized in the lateral coordinates parallel 
to the surface and in the rotational degrees of freedom of the molecule.
The associated modes become ``frustrated'' which leads to zero-point 
energies in these degrees of freedom. This is confirmed by Tab.~\ref{zeros} 
where the zero-point energies of the hydrogen molecule at the
minimum barrier position are collected.
Note that the zero-point energies in the frustrated rotational 
and lateral modes are still small, in particular the rotational modes.
This is caused by the fact that the molecular bond is essentially 
not elongated at the minimum barrier position so that the molecular 
interaction with the surface is still rather isotropic.

However, the sum of all zero-point energies at the minimum barrier position
is in fact 0.08~eV larger than the gas-phase zero-point energy of
hydrogen which is just half the molecular vibrational frequency
$\frac{1}{2} \hbar \omega_{\rm gas} =$~0.258~eV. The combined
effect of all zero-point energies at the minimum barrier position
is to enlarge the effective barrier height by 0.08~eV to 
$E_b^{eff} = $~0.17~eV. The absence of zero-point effects in the classical 
dynamics causes the strongly enhanced sticking probability in the classical
calculations. Still there is already a significant quantum sticking
probability for kinetic energies below this effective barrier height.
A detailed dynamical analysis suggests that this is due to the
combined effect of steering and tunneling \cite{Gro98pub}. Classically 
steering does not help to traverse the barrier region for energies below
the barrier height. Also for energies only slightly larger than
the barrier height steering is not very efficient. This is different
in quantum dynamics where steering can already occur in the tunneling
regime. Therefore it is very effective for energies close to the
minimum barrier height with respect to the classical results.

\begin{figure}[t]
\unitlength1cm
\begin{center}
   \begin{picture}(10,6.5)
\put(-1.,-5.){   {\epsfysize=18.cm  
          \epsffile{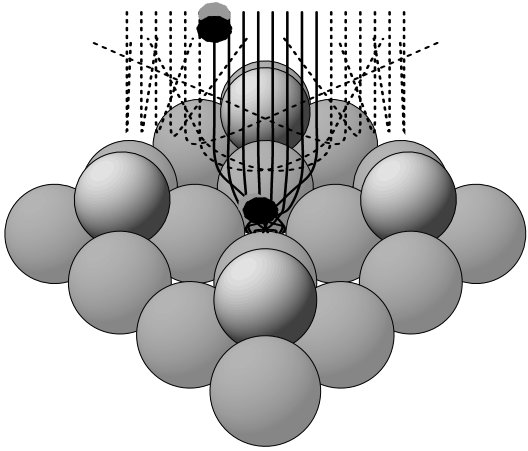}} }
   \end{picture}

\end{center}
   \caption{Illustration of steering at the (2$\times$2) sulfur covered
            Pd(100) surface. Center-of-mass trajectories of H$_2$ molecules 
            are shown impinging with an kinetic energy of
            $E_i = 0.3$~eV on the diagonal of the
            (2$\times$2) unit cell. Dashed lines correspond to
            reflection events while the full lines depict adsorption
            events.}

\label{h2spd_steer}
\end{figure}

Although the dissociation of hydrogen on the sulfur covered Pd(100)
surface in contrast to the clean surface is hindered by barriers, there 
is also significant steering of the impinging molecules to low-barrier
configurations operative. This is reflected by the fact that the
calculated sticking probabilities are
much larger than what one would expect from the hole model \cite{Kar87}
(see Fig.~\ref{stick}). This steering has been confirmed by
analysing swarms of trajectories and is illustrated in Fig.~\ref{h2spd_steer}.
The center-of-mass trajectories of H$_2$ molecules impinging on 
the diagonal of the unit cell of the (2$\times$2) sulfur covered Pd(100)
surface are shown. The initial kinetic energy is 0.3~eV.
The molecular axis is parallel to the surface, but the initial 
azimuthal orientation is chosen in such a way that in the sudden 
approximation all initial conditions do not lead to dissociative
adsorption since all incoming molecules hit barriers larger than 0.3~eV.
The strongly repulsive potential directly above the sulfur atoms 
extends rather far into the gas-phase \cite{Wei97}. Incoming molecules
are thus diverted from this high-barrier sites. Near to the sulfur atoms
the molecules are still scattered back into the gas-phase, but closer
to the center of the (2$\times$2) unit cell the molecules are focused
to the minimum barrier site. This focussing is accompanied by a
reorientation of the molecular axis so that the molecule dissociates
in the configuration depicted by the inset of Fig.~\ref{elbow}a.

\begin{figure}[t]
\unitlength1cm
\begin{center}
   \begin{picture}(10,6.5)
\centerline{   \rotate[r]{\epsfysize=8.cm  
          \epsffile{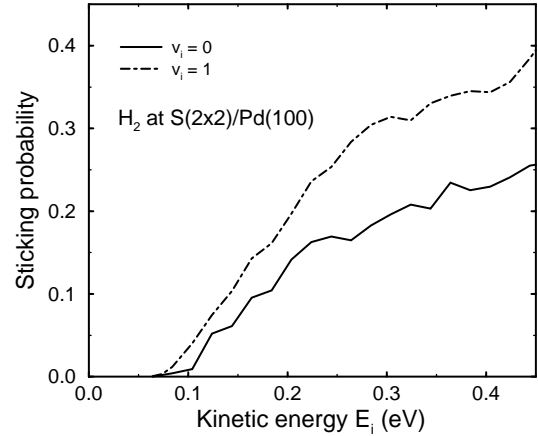}} }
   \end{picture}

\end{center}
   \caption{Dependence of the quantum sticking probability versus 
ikinetic energy for a hydrogen beam under normal incidence at a 
S(2$\times$2)/Pd(100) surface on the initial vibrational state $v$.
Solid line: $v_i = 0$, dash-dotted line: $v_i = 1$. The molecules
are initially in the rotational ground state.}
\label{state_stick}
\end{figure}

The effect of initial vibrational motion on the dissociation
process is demonstrated in Fig.~\ref{state_stick} where state-specific
sticking probabilities as a function of the incident kinetic energy are 
plotted. As Fig.~\ref{elbow}a shows, the minimum barrier to dissociation 
is at a position where the bond-length of the molecule is still
not significantly elongated. In such a situation it is usually anticipated
that the vibrational and translational degrees of freedom are almost
uncoupled so that vibrational energy of the impinging molecules cannot
be used to overcome the barrier (see, e.g., Ref.~\cite{Dar95rep}).
Consequently, it has been predicted \cite{Wil96S} that the sticking 
probability of H$_2$ at S(2$\times$2)/Pd(100) should show no strong 
dependence on the initial vibrational state of the molecule. 
This prediction corresponds to the so-called Polanyi rules which have been
formulated for gas-phase dynamics thirty years ago \cite{Pol69}.
Apparently the rules have to be modified for strongly corrugated surfaces.
Figure~\ref{state_stick} demonstrates that initial vibrational excitation 
leads to a significant increase in the sticking probability in this
early barrier system. As Fig.~\ref{elbow}b reveals,
there are molecular configurations for which the bond is significantly
extended at the barrier position. Adsorbing molecules do not only probe 
the minimum barrier of the PES, but also other regions of the PES where 
vibrational energy can be efficiently used to overcome the dissociation 
barriers. This result of the vibrationally enhanced dissociation 
is actually in agreement with the experimentally observed vibrational 
over-population in thermal hydrogen desorption from sulfur 
covered Pd(100) \cite{SchDiss91}
invoking the principle of microscopic reversibility.

It is interesting to note that the quantum sticking probability
for molecules in the first excited vibrational state is close
to the classical sticking probability for non-vibrating molecules
(see. Fig.~\ref{stick}). For the hydrogen dissociation on the
clean Pd(100) surface is has been shown that zero-point effects
can cancel out of the dynamics calculations if the sum of
all zero-point energies remains constant along the dissociation
path \cite{Gro97Vac}. For H$_2$ at S(2$\times$2)/Pd(100) this sum 
increases as Tab.~\ref{zeros} shows. Now the H-H vibration
corresponds to the fastest mode in the dissociation dynamics,
and a comparison of 5D vibrationally adiabatic with full 6D
calculations has shown that the vibrations indeed follow the
change of the vibrational frequency during the dissociation
almost adiabatically \cite{Gro96CPLa}. For molecules in the first excited
vibrational state the transfer from vibrational energy to
translational energy due to the lowering of the vibrational
frequences can compensate the increase in the zero-point energies
in the rotational and lateral modes leading to the
described cancellation effect.

\section{Hydrogen interaction with semiconductor surfaces}

The {\it ab initio} studies of hydrogen dissociation on metal surfaces
provide a rather complete picture of the dissociation process. The 
quantitative agreement can still be improved by, e.g., the inclusion
of dissipation processes to substrate phonons or electron-hole pairs,
the general mechanisms, however, seem to be understood (of course
one should always be careful with such optimistic
statements). For the interaction of hydrogen with semiconductor
surfaces, on the contrary, there is still no general agreement
on the fundamental mechanism, especially in the well studied system 
H$_2$/Si. The reason lies in the fact that the rearrangement
of the semiconductor surface upon hydrogen adsorption can no longer be 
neglected as for metal systems. 
This makes the adsorption/desorption dynamics and their
theoretical description more complicated.

In the following I will first discuss the dissociation dynamics
of hydrogen on Si(100). Then I will address a recent {\it ab initio} 
study of the reaction of atomic hydrogen with the hydrogen-passivated
Si(100) surface.

\subsection{Dissociation on semiconductor surfaces}

The system H$_2$/Si plays the same role for the hydrogen dissociation
on semiconductors that H$_2$/Cu has for metal surfaces: it serves as
a model system for which an abundance of experimental and theoretical
studies exist (see for example the recent review by Kolasinski \cite{Kol95}).
Besides, this system is also of great technological relevance for, e.g.,
the etching and passivation of Si surfaces or the growth of Si crystals.

\begin{figure*}[t]
\unitlength1cm
\begin{center}
   \begin{picture}(10,6.0)
      \includegraphics{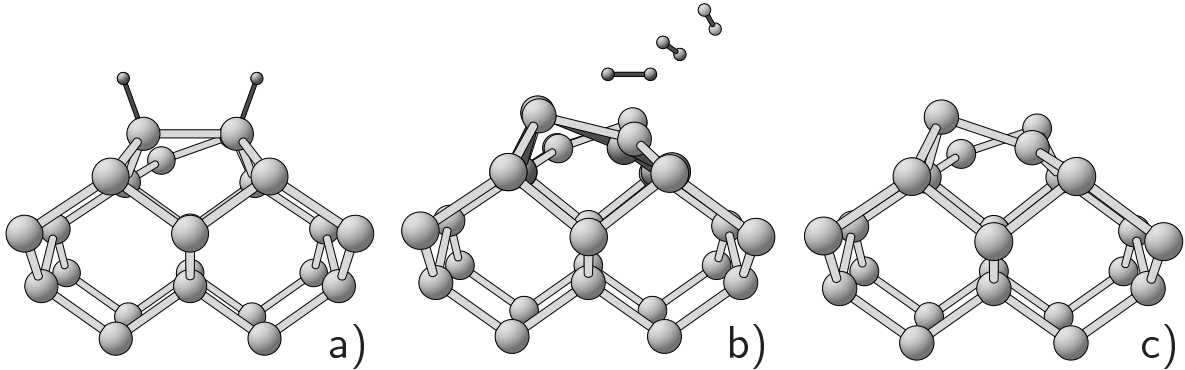}
   \end{picture}
\end{center}
\caption{a) Hydrogen covered Si\,(100) surface (monohydride).
b) Snapshots of a trajectory of D$_2$ desorbing from Si\,(100)
starting at the transition state with the Si atoms initial at rest
\protect\cite{Gro97H2Si}. 
The dark Si atoms correspond to the
Si positions after the desorption event. 
c) Clean anti-buckled Si\,(100) surface.
\label{figdes}}
\end{figure*}

Desorption of hydrogen from Si(100) shows first-order kinetics
\cite{Sin89,Hoef92,Flo93,Sha94}. For associative desorption one normally 
expects second-order kinetics since two atoms have to find each other on
the surface before they can desorb. The unusual first-order desorption
kinetics has been explained by a prepairing 
mechanism \cite{Doren96,Hoef92,Flo93}: Desorbing molecules originate
from the same dimer since it is energetically favorable
for two hydrogen atoms to bind on the same dimer rather than on
two independent dimers (see Fig.~\ref{figdes}a).
 
The so-called barrier puzzle has further fueled the interest for this system:
While the sticking coefficient of molecular hydrogen on Si surfaces
is very small \cite{Lie90,Kol94JCP,Bra95,Bra96CPL}
indicating a high barrier to adsorption, in a laser-induced desorption 
experiment an almost thermal mean kinetic energy  of the molecules was found 
\cite{Kol94PRL} indicating a low adsorption barrier. In order to 
explain this puzzle it was suggested to take the strong surface rearrangement 
of Si upon hydrogen adsorption into account \cite{Kol94PRL}: The hydrogen 
molecules impinging on the Si substrate from the gas phase typically
encounter a Si configuration which is unfavorable for dissociation, 
while desorbing hydrogen molecules leave the surface from a rearranged 
Si configuration with a low barrier.
In the adsorption process such a surface rearrangement can only be achieved
by thermal excitations of the lattice since due to the mass mismatch the
silicon substrate atoms are too inert to change their configuration
during the time the hydrogen molecule interacts with the surface.
The strong surface rearrangement of Si(100) upon hydrogen 
adsorption/desorption is illustrated in Fig.~\ref{figdes}.

With the assumption of a lattice rearrangement energy of about 0.7~eV 
existing experimental adsorption and desorption results could be 
reproduced by quantum dynamical model calculations 
\cite{Bre94a,Bra96PRB,Bre97}.
The predicted strong dependence of the adsorption probability
on the surface temperature was later confirmed experimentally \cite{Bra95}.

So far the dissociation mechanism seemed to be at least qualitatively 
understood. However, there are some disturbing experimental uncertainties.
Measurements of the dissociative adsorption probability of H$_2$/Si(100) 
differ by almost three orders of magnitude 
from each other~\cite{Kol94JCP,Bra96CPL}. 
In order to explain the large difference it had been
speculated~\cite{Bra96CPL}, e.g., that the Si sample used for the adsorption 
study of Ref.~\cite{Kol94JCP} exhibits an unusual low barrier due to its
high dopand concentration of $n \approx 10^{19}$~cm$^{-3}$.

\begin{figure*}[th]
\unitlength1cm
\begin{center}
   \begin{picture}(18,12.0)
\put(-1.,0.){ {\epsfxsize=10.cm  
          \epsffile{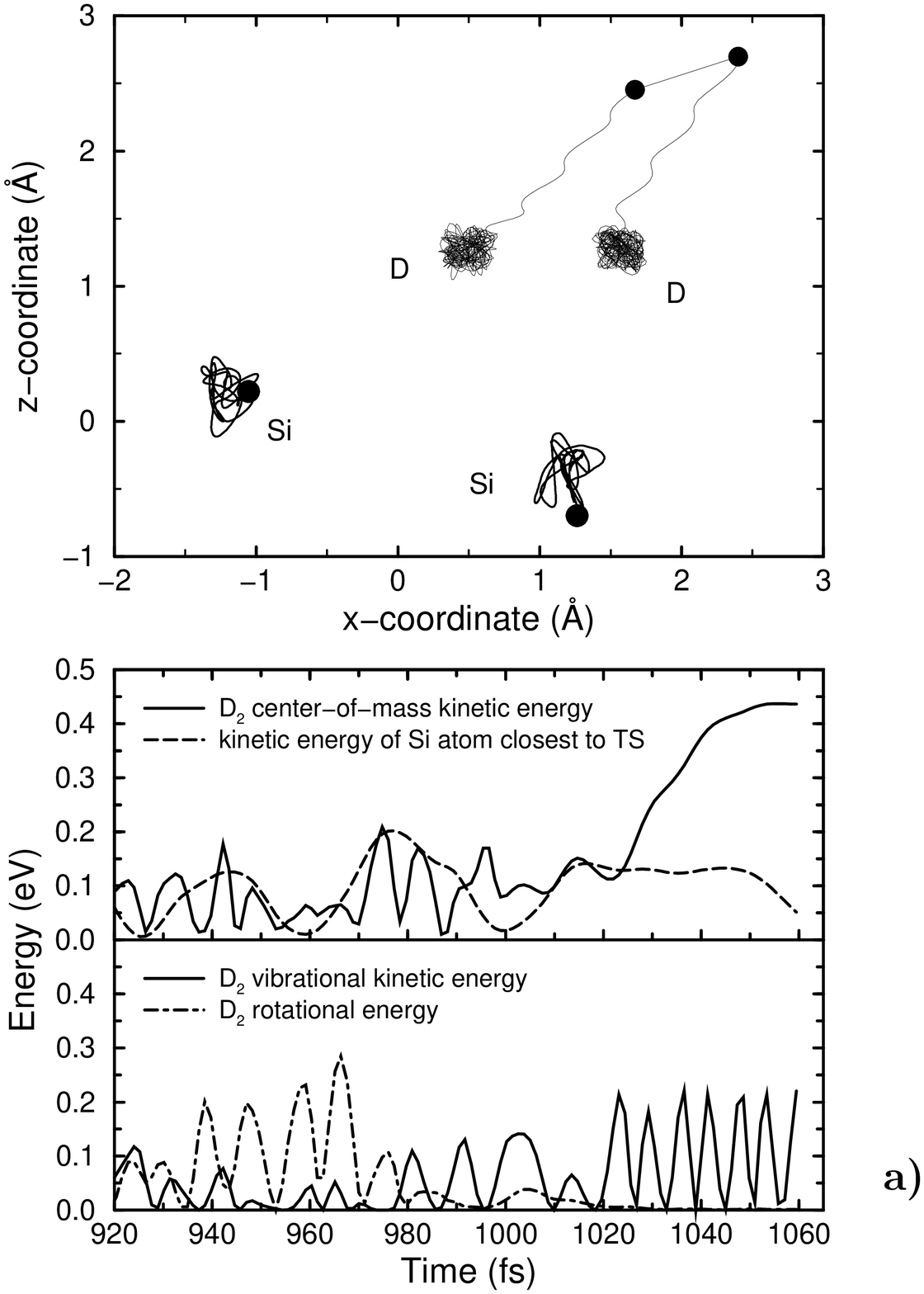}} }
\put(8.2,0.){ {\epsfxsize=10.cm  
          \epsffile{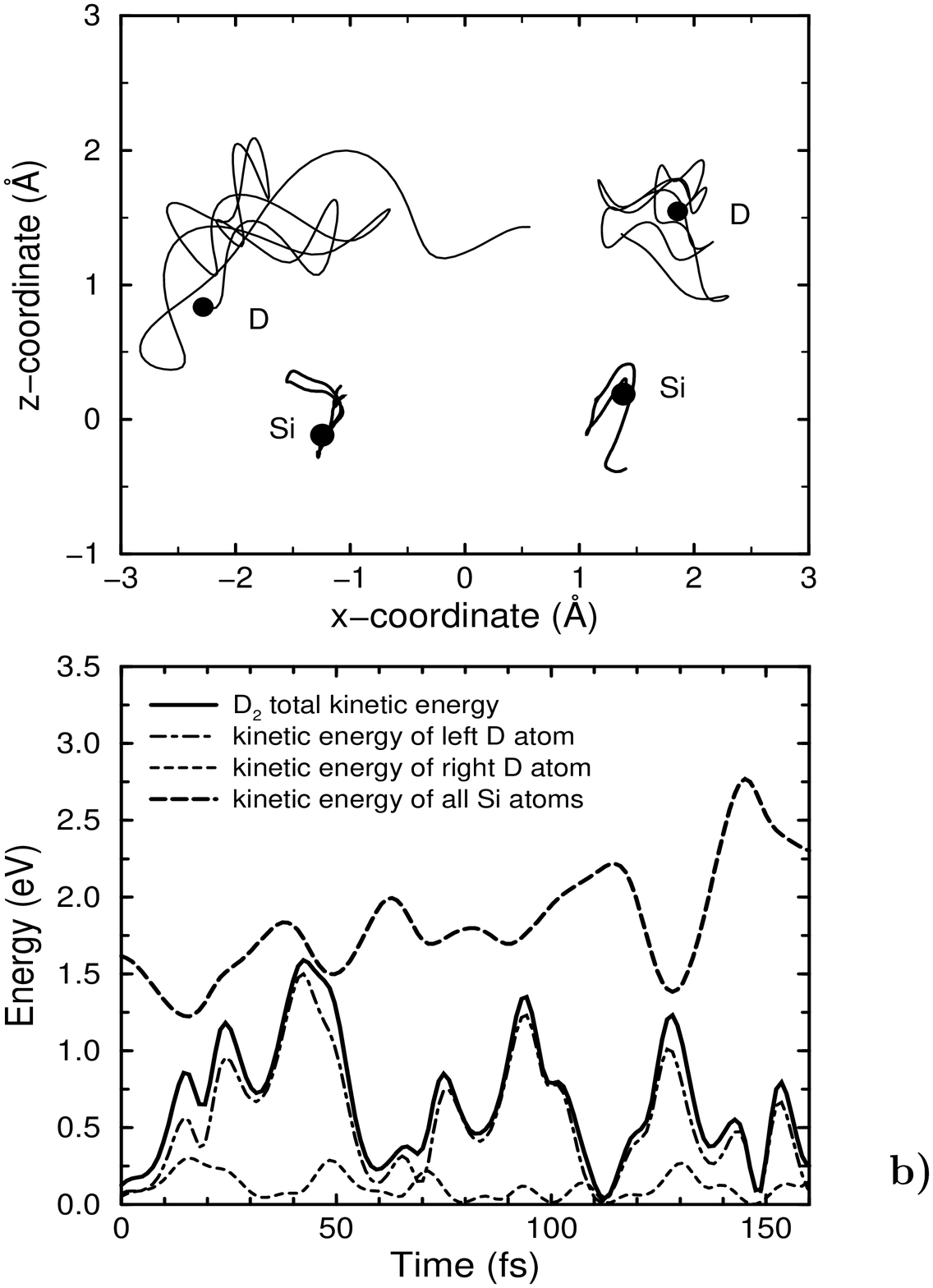}} }

   \end{picture}
\end{center}
   \caption{{\it Ab initio} molecular dynamics trajectory of D$_2$
            desorbing from Si(100). The initial conditions correspond
            to a surface temperature of  $T_s = 920$~K.
   a) Thermal desorption trajectory started close to the transition 
   state (TS). 
   Upper panel: $x$ and $z$
   coordinates of the trajectories of the two deuterium atoms and the
   two Si atoms of the dimer underneath the TS.
   The final positions of the atoms are marked by the large dots.
   Lower panel: energy redistribution as a function of the last 140~fs of
   the run-time of the trajectory. The different energy channels are
   described in the legend of the figure.   
   b) Time-reversed trajectory of the desorption event started at the TS. 
   The upper and lower panels correspond to the ones in a).
 }
\label{thermtraj}
\end{figure*}

From the theoretical point of view, the situation is even more confusing.
The exact adsorption and desorption mechanism is strongly debated. 
Total-energy calculations using the cluster approach 
\cite{Wu93,Jin93,Nac94,Nac95,Jin95,Rad96} have found activation barriers for 
associative desorption of hydrogen from Si(100) which are roughly 1~eV larger 
than the experimentally found value of about 2.5~eV \cite{Hoef92,Flo93}. 
Based on these findings, the prepairing mechanism has been disregarded,
and defect-mediated desorption mechanisms had been proposed 
\cite{Wu93,Nac94,Nac95,Jin95,Rad96,Rad96PRB,Rad97PRB}. These mechanisms, 
however, have difficulties explaining the observed first-order desorption
kinetics and the measured high prefactor of the desorption rate
(for a thorough discussion of these calculations, see the review
by Doren \cite{Doren96}). The question arises whether a cluster
really represents a good model for an infinite substrate. For example,
in most of the cluster calculations the Si(100) surface was found to
reconstruct with a symmetric dimer structure \cite{Nac94,Rad96PRB,Jin92PRB}
while in slab calculations the anti-symmetric buckled dimer structure 
is the lowest energy configuration \cite{Dab92,Nor93} 
which is also well-established experimentally \cite{Wol92,Bul95,Smith96}. 
It was suggested that the relaxation
constraints used in the cluster studies do not model the Si(100) surface
correctly \cite{Yang97}. Furthermore, the reported barriers for
the {\em same} process differ by almost 1~eV between three different
sets of cluster calculations \cite{Wu93,Jin93,Nac94} indicating
the difficulty of such calculations.

On the other hand, in detailed DFT-GGA  calculations of the H$_2$/Si\,(100)
potential energy surface (PES) using the supercell 
approach~\cite{Kra94,Kra96PRB,Peh95,Vit95} and also in clusters
calculations \cite{Pai95} desorption barriers for hydrogen atoms
originating from the same Si dimer were obtained that 
are in agreement with the experimentally determined values.
This pre-pairing mechanism, as mentioned above, 
is consistent with the observed first-order kinetics in
experiment \cite{Sin89,Hoef92,Flo93,Sha94}.
The adsorption barriers were found to be about 0.3~-~0.4~eV with 
the substrate rearrangement energy being roughly 0.15~eV. 
These barrier heights are also controversely discussed. It was
argued that the GGA-PW91 functional used in the DFT studies underestimates 
the H$_2$ elimination barriers from silanes, i.e., small 
molecules \cite{Nac96}. It is, however, questionable whether the
results for small molecules are directly transferable to
surface systems.

Subsequently, based on the information from the slab 
calculations \cite{Kra94,Kra96PRB} two different three-dimensional quantum 
dynamical studies were performed \cite{Kra96,Lun96}, where the relaxation 
of the Si substrate upon hydrogen adsorption was approximated by one
idealized Si phonon coordinate. Although both 
calculations used the same {\it ab initio} energies as a source, the 
results of the dynamical calculations did not agree quantitatively.
However, in both studies the dynamical coupling of the desorption path to 
the Si substrate was low, i.e., the desorbing molecules had more
excess translational energy than found in the experiment~\cite{Kol94PRL}.

These different uncertainties together with the observed unusual 
first-order desorption kinetics make the H$_2$/Si(100) system to one of the
most controversely discussed systems in the field of gas-surface dynamics.
Since barrier calculations alone will not settle the controversy about
the desorption mechanism \cite{Doren96},  dynamics calculations
are necessary in order to find out how the potential energy
at the barrier is distributed over the various degrees of freedom
of this system (hydrogen vibration, rotation, and translational energy,
and vibrations of the Si substrate). Due to the importance
of lattice relaxation effects on the desorption dynamics, substrate
degrees of freedom have to be taken into account in the dynamics
simulations. This makes a full quantum dynamical treatment not possible
at the moment. {\it Ab initio} molecular dynamics calculations using 
both the supercell approach \cite{Gro97H2Si}
and clusters \cite{Silva97} have therefore been performed. In the supercell
calculations DFT-GGA was employed to determine the desorption dynamics
of deuterium desorbing from the monohydride (MH) state using
a (2$\times$2) surface unit cell and a five-layer Si slab \cite{Gro97H2Si}
(see Fig.~\ref{figdes}).
In the cluster calculations hydrogen desorption was treated on the
complete active space self-consistent field (CASSCF) level.
Both desorption from a single monohydride using a 
Si$_9$$\bar{\rm H}_{12}$H$_2$ cluster and from an isolated dihydride (ID)
using a Si$_{10}$$\bar{\rm H}_{14}$H$_2$ cluster
were studied \cite{Silva97}, where $\bar{\rm H}$ denotes the modified
hydrogen atoms terminating the cluster.

Since the barrier to associative desorption of hydrogen from Si(100) is 
rather high, in both sets of {\it ab initio} molecular dynamics 
calculations \cite{Gro97H2Si,Silva97} the trajectories were not started
with the hydrogen at the atomic adsorption positions. Instead, the
molecular dynamics calculations were initialized with the hydrogen atoms
close to the transition state with initial velocity distributions
that correspond to the surface temperatures used in the experiments.
In the supercell calculations, additionally the time-reversed trajectories
starting at the transition state were determined in order to make sure that 
all trajectories correspond indeed to desorption and not to scattering events
\cite{Gro97H2Si}.

In the DFT-GGA calculations additional trajectories were determined 
with the Si lattice initially  at rest, i.e., at a surface temperature 
of $T_s = 0$~K. Snapshots of such a calculated trajectory are shown in 
Fig.~\ref{figdes}b). The dark Si atoms
correspond to the relaxation of the Si lattice after the desorption event.
Approximately 0.1~eV of the potential energy at the transition state
is transferred to vibrations of the Si lattice which is a rather large
amount compared to hydrogen/metal systems. 
At the transition state the interatomic hydrogen distance is about 
40{\%} larger than the gas-phase bond length; 
consequently molecular vibrations are excited during the desorption,
as is well known for a long time in associative desorption studies
(see, e.g., ref.~\cite{Dar95rep}).

Similar results have been found in the DFT-GGA calculations
for initial conditions corresponding to a surface temperature
of  $T_s = 920$~K. In total 34 ``thermal'' desorption trajectories 
were calculated. 
Figure~\ref{thermtraj}a) shows an example of such a trajectory.
First the surface was equilibrated for approximately
1~ps with the deuterium atoms kept close to the transition
state by an auxiliary potential. Then  
the extra potential was switched off, the molecule was allowed to
desorb and the distribution of the energy into the various degrees
of freeedom was monitored.

The projection of the trajectories of the desorbing deuterium 
molecule and of the Si dimer closest to the transition state onto the 
$xz$-plane is shown in Fig.~\ref{thermtraj}a). Due to the asymmetry
of the transition state the molecule does not desorb in the direction 
normal to the surface. The oscillations in the trajectories
correspond to the vibrational excitation of the desorbing D$_2$.
In addition, the energy redistribution during
the last 140~fs of the run is plotted where, e.g., the vibrational 
excitation can be clearly followed. The quenching of
the rotational motion is evident which is caused by the strong
anisotropy of the PES. Furthermore, the deuterium molecule 
is strongly accelerated during the desorption event and ends up with
a kinetic energy of roughly 0.4~eV which is larger than
the value of $2k_B T_s = 0.16$~eV expected for thermal equilibrium.

The configuration at which the extra potential was switched off
was also used to start a time-reversed trajectory. As mentioned
above, the main reason was to check whether the trajectory 
describes a true reactive event. On the other hand, the time-reversed
trajectory simultaneously corresponds to a dissociative adsorption
event and allows to follow the energy distribution during the 
adsorption. Once the molecule has passed the transition state,
the single atoms are strongly accelerated towards their atomic
adsorption positions; together they can gain an energy of 2.5~eV.
Figure~\ref{thermtraj}b) shows that indeed the two D atoms first
become very fast, in particular the left deuterium atom that has
the longer distance to its adsorption position. Within 50~fs the
D atoms have gained about 1.5~eV in kinetic energy. Although the
mass mismatch between the deuterium and the silicon atoms is so large,
there is still a significant energy transfer to the silicon lattice
due to the high kinetic energy of the D~atoms. They bounce into the 
lattice so violently that within 150~fs 1~eV of the kinetic
energy is transferred to the silicon atoms, as Fig.~\ref{thermtraj}b)
reveals. Furthermore, Fig.~\ref{thermtraj} nicely demonstrates how
the atoms of the Si dimer are moving towards a symmetric configuration
in the case of the D$_2$ dissociative adsorption, while in the
case of the associative desorption the buckling angle is further
increased with respect to the transition state configuration.

\begin{table*}[t]
\begin{center}
\begin{tabular}{|l|c|c|c|c|c|}
\hline
  &  $\langle E_{tot} \rangle$ & $\langle E_{kin} \rangle$ & 
 $\langle E_{vib} \rangle$ & $\langle E_{rot} \rangle$ 
&  $\langle \theta_f \rangle $   \\ 
\hline 
D$_2$, MH, DFT-GGA  & $0.70  \pm 0.18$ & $0.55  \pm 0.14$  
 & $0.12  \pm 0.09$  &  $0.03 \pm 0.05$ & $39.6^{\circ} \pm 10.1^{\circ}$ \\  
H$_2$, MH, CASSCF &  $0.72  \pm 0.20$ &  $0.6  \pm 0.1$  
 & $0.09  \pm 0.08$  &   $0.03  \pm 0.02$ & $56^{\circ} \pm 7^{\circ}$ \\  
H$_2$, ID, CASSCF
 &  $0.94  \pm 0.20$ &  $0.7  \pm 0.1$  
 & $0.2  \pm 0.1$  &   $0.04  \pm 0.03$  & $43^{\circ} \pm 17^{\circ}$ \\  
D$_2$, Experiment & $0.34 \pm 0.05$ &  $0.17 \pm 0.03$ & $0.15 \pm 0.03$      
 & $0.03 \pm 0.01$ & $28^{\circ} - 31^{\circ}$\\
\hline
\end{tabular}
\end{center}
\caption{Mean energy distribution and final desorption angle of hydrogen 
molecules desorbing from a Si(100) surface. The results of the DFT-GGA 
calculations are averaged over 34 desorption 
trajectories corresponding to D$_2$ desorption
from the monohydride (MH) phase  at a surface temperature 
of $T_s = 920$~K ($k_B T_s = 0.079$~eV) \protect\cite{Gro97H2Si}.
The cluster calculations at the CASSCF level for H$_2$ desorption from a MH and
an isolated dihydride (ID) are averaged over ten trajectories for each case
at a surface temperature of $T_s = 780$~K ($k_B T_s = 0.067$~eV)
\protect\cite{Silva97}.  The experimental results are 
collected from Ref.~\protect\cite{Kol95}.
The translational energy has been been determined in a laser-induced
desorption experiment where the surface temperature was estimated
to be $T_s = 920$~K \protect\cite{Kol94PRL}
while the vibrational and rotational energy have been measured at a
surface temperature of $T_s = 780$~K \protect\cite{Kol92}.
The mean vibrational energy corresponds to the classical value using 
the vibrational temperature measured in the experiment (see text). 
The experimental mean desorption angles are derived from 
Ref.~\protect\cite{Park93}, where the angular distribution in desorption
has been measured for two different hydrogen coverages.
\label{tabdes}}

\end{table*}

The mean total, kinetic, vibrational, 
and rotational energies and final angles of the hydrogen molecules
desorbing from Si(100) obtained from the {\it ab initio} molecular
dynamics studies are compared to the experimental results in
Tab.~\ref{tabdes}. Since there is apparently some confusion
about doing a reasonable comparison between theory and experiment,
I will briefly expound on the given values. The vibrational and
rotational population of hydrogen molecules desorbing from Si(100)
was measured state-specifically using resonance-enhanced
multiphonon ionization detection \cite{Kol92}. This gives information
about the population of the quantum mechanical eigenstates of the
hydrogen molecule. For example, it has been measured that
at a surface temperature of $T_s = 780$~K 1\% of desorbing
H$_2$ molecules are in the first vibrational eigenstate \cite{Kol92}. This
corresponds to a fraction that is 20 times larger than expected
in thermal equilibrium and is described by the term ``vibrational
heating''. In Ref.~\cite{Silva97} this population was used to determine
the mean vibrational energy of desorbing H$_2$ molecules. They obtained
a value of 0.264~eV. However, 98\% of this value comes from the
zero-point energy of the ground-state molecules, only 2\% from molecules 
in the first excited vibrational state. Had the experiments not given
any vibrational heating, this mean vibrational energy would have hardly 
been changed at all. 
In the {\it ab initio} molecular dynamics calculations the
dynamics of the hydrogen nuclei is treated fully classically, there are
no zero-point energies considered. Since the H$_2$ zero-point energy
is so large (three times larger than the thermal energies in the
experiment) it makes no sense comparing a value derived from a quantum 
mechanical analysis with a classically obtained value. Therefore
I have put in Tab.~\ref{tabdes} for the experimental mean vibrational energy
the value $\langle E_{vib} \rangle^{exp} =  k_B T_{vib}$, 
where $T_{vib} = 1700$~K
was taken from the measured population \cite{Kol92}. This procedure is
also not ideal, but certainly closer to a classical analysis.
A real quantitative comparison between experiment and theory can only
be done for a quantum treatment of the hydrogen vibrational dynamics.
In the three-dimensional quantum dynamical studies, e.g, the correct
experimental population of vibrationally excited states in desorption
is reproduced \cite{Bra96PRB,Bre97,Kra96}.
These problems do not arise for the mean translational and rotational
energies. Since these modes are free in the gas-phase, there are no
zero-point energies associated with them.

Furthermore, in Ref.~\cite{Silva97} a mean experimental kinetic energy 
of 0.34~eV for H$_2$ desorption was derived which was said to be scaled
from the experimental value of 0.17~eV for D$_2$. This scaling seems
very questionable to me. H$_2$ and D$_2$ have exactly the same interaction 
potential. The gain in kinetic energy is determined by the change in
the potential energy which is the 
same for H$_2$ and D$_2$. And classically different isotopes follow
exactly the same trajectories as a function of the kinetic energy
if no energy transfer to substrate degrees of freedom is considered
\cite{Gro98PRB}. Hence H$_2$ and D$_2$ should have approximately the
same kinetic energy in desorption. There can be an isotope effect
due to quantum mechanical effects or due to energy transfer to the
silicon lattice. However, the sign and magnitude of this isotope
effect is not clear {\it a priori}; it has to be determined by
detailed calculations. Certainly it does not follow
such a simple scaling law as assumed in Ref.~\cite{Silva97}.

Comparing the theoretical and experimental data in Tab.~\ref{tabdes},
it should furthermore be noted that the reported theoretical values do not
really correspond to any thermal average. For that purpose the number
of 34 and 10 calculated trajectories, respectively, is much too low.
Hence the theoretical results have to be interpreted cautiously.
Still they should be able to indicate the general trends in the
energy distribution of desorbing molecules since in thermal
desorption the molecules mainly originate from a small fraction
of the configuration space close to the transition state
(the so-called keyhole effect \cite{Gro95JCP}). As for the mean
desorption angle, note that this value correspond to an average
over all azimuthal angles with $0 \leq \theta_f < 90^{\circ}$.
Hence even if the angular distribution for fixed azimuth is 
peaked in the direction normal to the surface, $\langle \theta_f \rangle $ 
will still be larger than zero.

It is obvious that all simulations reproduce the vibrational heating, 
i.e. \mbox{$\langle E_{vib} \rangle > k_B T_s$}, and the rotational cooling, 
i.e. \mbox{$\langle E_{rot} \rangle < k_B T_s$}. 
The final desorption angles are slightly larger than the experimental
result \cite{Park93}. The mean kinetic energy, however, is much larger 
for all {\it ab initio} molecular dynamics calculations than the experimental 
value for D$_2$ of 
$\langle E_{kin} \rangle^{exp} = 0.17$~eV~$\gtrsim 2k_BT_s$~\cite{Kol94PRL}. 
In the cluster calculations, molecules desorbing from the isolated
dihydride are even faster than molecules desorbing from the monohydride.
It is remarkable that the results of the supercell and the cluster 
calculations are rather similiar for the mean final energies
of hydrogen molecules desorbing from the monohydride state.

The theoretical mean desorption angles are somewhat larger than the
experimental values. However, due to the low number of trajectories
obtained in the calculations the discrepancies are not conclusive to rule
out any particular mechanism.

In conclusion,
the analysis of the data of Tab.~\ref{tabdes} gives no reason to disregard
the prepairing mechanism with respect to any defect-mediated mechanism.
According to the results of the {\it ab initio} molecular dynamics 
runs \cite{Silva97}, the isolated dihydride mechanism does even worse
compared to the experiment than does the prepairing mechanism. And since
the observed first-order kinetics in hydrogen desorption from Si(100)
follows very naturally from the prepairing process, this still seems
to be the most probable mechanism of desorption. Just recently there has 
been a very detailed temperature programmed desorption study of
hydrogen desorption from Si(100) at low coverages which gave no
indication that defect sites are involved in the desorption 
process \cite{Flo98}. Similar conclusions can be drawn from another
recent study about hydrogen adsorption on Si(100) which
allowed to discriminate between hydrogen adsorption on step and defect
sites and on terrace sites \cite{Rat98}.

The question still remains why the kinetic energy in desorption
comes out much larger in the {\it ab initio} molecular dynamics
runs than in the experiment. Uncertainties due to the bad statistics
of the molecular dynamics runs
or quantum dynamical effects are not believed to be responsible
for the large difference \cite{Gro97H2Si,Lun96}.
In all the {\it ab initio} calculations the electrons are assumed
to follow the motion of the nuclei adiabatically in their ground
state. It has therefore been speculated that the excess energy
at the transition state might be transferred to surface electronic
excitations \cite{Gro97H2Si,Doren96} thus taking energy away from
the molecular degrees of freedom.
Besides, the laser-induced desorption used in the experiment
to determine the translational energy in desorption \cite{Kol94PRL}
might not correspond to thermal desorption events described by the 
{\it ab initio} molecular dynamics runs. Recent DFT calculations indicate
that laser melting of bulk silicon leads to a transition to a
metallic liquid state~\cite{Sil96} which could also occur during 
the laser-induced desorption used in the experiment. 
Furthermore, at surface temperatures above 900~K the clean Si(100)
surface undergoes a semiconductor-metal transition \cite{Gav96}.
At a metallic surface the barrier distribution might be different.
In addition, the excitation of electron-hole pairs is much more
probable leading to an effective dissipation channel for any
excess energy at the transition state. Since all these considerations
about the correct desorption mechanism rely on {\em one} laser-induced
desorption experiment, it is highly desirable that the mean kinetic
energy of hydrogen molecules desorbing from Si(100) is measured
independently by preferentially a thermal desorption experiment.

\subsection{Reactions of atomic hydrogen with hydrogen-passivated
semiconductor surfaces}

\begin{figure}[tb]
\unitlength1cm
\begin{center}
   \begin{picture}(10,5.5)
\centerline{   \rotate[r]{\epsfysize=9.5cm  
          \epsffile{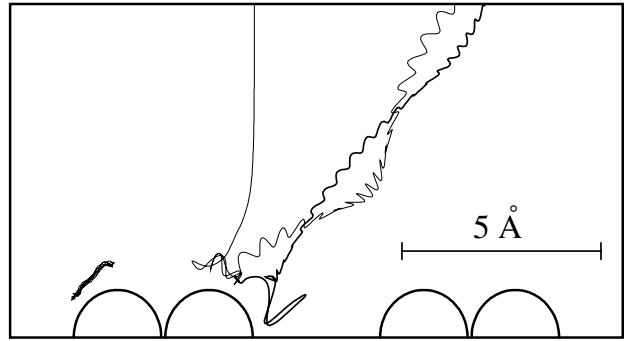}} }

   \end{picture}
\end{center}
   \caption{Example of a reactive trajectory for an H atom abstracting
            a D atom adsorbed on Si(100). The incident kinetic energy
            of the H atom is 0.2~eV (from Ref.~\protect\cite{Kra97}). }
\label{H_D_abst}
\end{figure}

The associative desorption reaction described in the previous section
is of the so-called Langmuir-Hinshelwood (LH) type in which both
reactants are equilibrated with the substrate before the reaction.
In the Eley-Rideal mechanism, on the other hand, an incoming gas phase
species is assumed to react directly with an adsorbate and abstract
it from the surface. In Fig.~\ref{H_D_abst} such an abstraction 
reaction is illustrated.
Since in the ER reaction mechanism only one atom-surface bond has
to be broken instead of two in the LH mechanism, the ER reactions
show a significant exothermicity leading to large translational
and internal energies in the desorbing products.
Although the concept has long been known, 
experimentally the Eley-Rideal mechanism has only recently been confirmed, 
at first for hydrogen abstraction reactions from metal surfaces
\cite{Kui91,Ret92ER,Ret96ER,Jach94}. In the meantime also hydrogen
abstraction reactions from Si surfaces have been observed \cite{Wid95,Bun98}.
These abstraction reactions have been found to be very efficient:
the reaction probability for abstraction on the saturated monohydride
surface is $0.5 \pm 0.1$ relative to the atomic adsorption probability
on the clean surface \cite{Wid95}. Since the adsorption probability
of hydrogen atoms on Si surfaces is of order unity \cite{Schu83},
this means that roughly half of the incoming molecules pick up
another hydrogen atom from the surface. However,
only a small fraction of the incoming atoms directly hit the
adsorbed hydrogen atoms. In order to explain the observed adsorption
kinetics and the high abstraction probability, it was therefore
suggested that also indirect processes take place \cite{Wid95}. 
The impinging atom does not immediately pick up another atom
from the surface but remains trapped at the surface while keeping most of
its energy before the abstraction reaction. This mechanism, which has
also been suggested for hydrogen abstraction reactions on metal surfaces
\cite{Weh98}, is called ``hot precursor'' \cite{Wid95} or 
``hot atom'' \cite{Kra97} mechanism 
and has first been addressed by Harris and Kasemo \cite{Har81}.

Theoretically ER reactions have first been treated by low-dimensional
quantum model studies in restricted geometries \cite{Kra91,Jack92,Per95}.
These studies focused on the strong vibrational excitation of the
molecules upon the abstraction reaction.
Such a low-dimensional treatment does not allow the description of the
hot atom process. In order to figure out the importance of these
indirect processes, it is still unavoidable to rather perform classical 
dynamical calculations with a high dimensionality instead of 
quantum calculations with a restricted dimensionality. However, this is a
reasonable approach since due to the large exothermicity of the reaction
quantum effects should be negligible. Classical trajectory calculations
of the reaction dynamics of atomic hydrogen with the hydrogenated
Si(100) surface have recently been performed by P. Kratzer \cite{Kra97,Kra98}.
In these studies the PES has been described in a modified LEPS form. 
The energy transfer to the Si substrate is taken into account by a
collective mode in the surface oscillator model.
In a first study, the LEPS parameters were derived from both experiment 
and {\it ab initio} calculations \cite{Kra97}. In a subsequent
treatment the PES parameters were all obtained by spin-polarized
DFT-GGA calculations in the supercell approach \cite{Kra98}.

Figure~\ref{H_D_abst} shows an example of a calculated reactive trajectory.
An incoming H atom is directly abstracting an adsorbed D atom. The fast
oscillations illustrate the strong vibrational excitation of the
product HD molecule while the slow oscillations correspond to the
rotational excitation. However, such a direct ER reaction is in fact
less probable than an indirect hot atom abstraction reaction, as the
{\it ab initio} molecular dynamics calculations reveal \cite{Kra98}.
For H atoms incident with kinetic energies above 1~eV less than
20\% of the produced HD molecules originate from direct ER reactions.

Briefly summarizing the theoretical results, it is found that the
abstraction probability, which agrees with the experiment within the
experimental uncertainty, decreases with increasing incident
kinetic energy. Most of the energy dissipated to the surface is
transferred to vibrations of the adsorbed hydrogens remaining on the surface
since the energy transfer to the Si substrate atoms is rather low
due to the large mass mismatch between H and Si. But the most important
channel of energy disposal in the reaction is the kinetic energy of
the hydrogen molecules. Just recently also the H absorption in and
adsorption on Cu(111) has been studied by molecular dynamics calculations
where the PES was derived from DFT calculations and expanded in
an effective-medium theory form \cite{Str98}. 
These calculations confirm that {\it ab initio}
molecular dynamics studies are an indispensable tool to complement
experimental information since contrary to the experiment the calculations 
allow a microscopic analysis of the reaction dynamics.

\section{Laser-induced desorption}

So far all reported {\it ab initio} dynamics studies employed the
Born-Oppenheimer approximation, i.e., the electrons were assumed to
follow the motion of the nuclei adiabatically. However, one very active 
field of experimental research during the last years has been the
study of desorption induced by electronic transitions, so-called
DIET processes \cite{Bre85}. These processes can either be induced
by electron impact (ESD) or by light (PSD). The abbreviations denote 
electron or photon stimulated desorption, respectively. In particular,
there have been numerous investigations of laser-induced desorption of small 
molecules from surfaces (for a review, see Ref.~\cite{Zim95}).

The theoretical description of the dynamics of desorption processes involving
electronically excited states is rather complex. In most studies the
electron dynamics is not taken into account explicitly, instead the electronic 
transitions are treated in the Frank-Condon or sudden 
approximation \cite{Gad87,Saal97}. In a quantum treatment of the dynamics
of the nuclei this corresponds to the jumping of wave packets between
different electronic states of the molecule-surface system. Usually only
the electronic ground state and one excited state are considered. In these
studies the total energy is not conserved, i.e., it is assumed that
the energy difference between the electronic states is entirely taken away
by the electrons. To overcome this problem, either the electron dynamics
have to be treated explicitly \cite{Har95} or the excitation spectrum of
the solid has to be taken into account \cite{Saal95}.

Although the modelling of laser-induced desorption can be
rather successful in elucidating dynamical aspects of the processes
(see, e.g., Refs.~\cite{Her95,Born97}), 
the potential curves for the electronic
excited states had to be guessed since there were no accurate calculations
for them. Furthermore, at metal surfaces the electronically excited states
are strongly coupled to the continuous electronic excitation spectrum
of the surface. It is questionable whether the description of a DIET
process with just one sharp
well-defined electronically excited state is reasonable in this 
strong-coupling regime.

\begin{figure}[tb]
\unitlength1cm
\begin{center}
   \begin{picture}(10,6.0)
\put(-2.,-10.){\epsfxsize=12.5cm  
          \epsffile{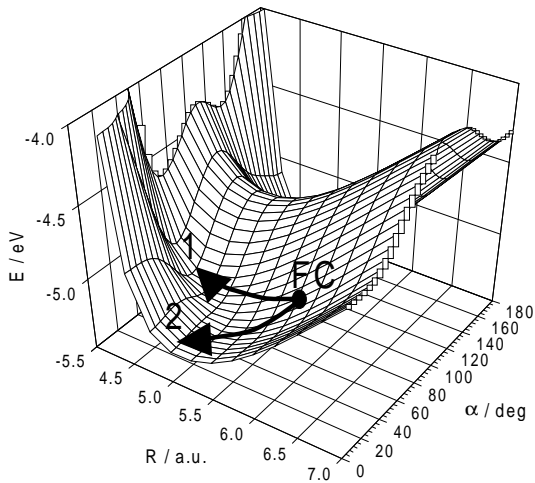} }

   \end{picture}
\end{center}
   \caption{Charge transfer PES of NO/NiO(100) as a function of the
            NO center of mass distance R from the surface and the
            polar orientation of the molecule $\alpha$. FC denotes
            the Franck-Condon point at which the wave packet propagation
            is started 
            (from Ref.~\protect\cite{Klue98PRL}). }
\label{NONiO_PES}
\end{figure}

This situation is different for the excited states at insulator surfaces
like oxides due to the large band gap. In addition, due to the strong 
ionicity of these systems a local description of the electronic structure
is reasonable. This is important insofar as density functional theory
is an electronic ground state theory and does not straightforwardly
allow the determination of excited state potentials (there are DFT 
calculations describing electronic excited states, though, 
see for example Ref.~\cite{Pan95}).  Hence one has to
employ quantum chemical {\it ab initio} methods which require the use
of a cluster approach. Indeed such an approach has been followed recently
for the first time. Kl{\"u}ner and coworkers have performed configuration 
interaction (CI) calculations for the construction of potential energy 
surfaces for electronically excited states describing the laser-induced
desorption of NO from NiO(100) \cite{Klue97,Klue98PRL,Klue96JCP}.
In the CI calculations a NiO$_5^{8-}$ cluster was embedded in a
semi-infinite Madelung potential of $\pm$2 point charges for the
simulation of the NiO(100) surface.
The calculated ground state potential did in fact not reproduce the
experimentally found binding energy of 0.52~eV \cite{Kuhl91}, confirming
the well-known problems in determining binding energies at surfaces
by cluster calculations \cite{Whi96}. The {\it ab initio} minimum has 
therefore been scaled to fit the experimental data \cite{Klue98PRL}.
One specific charge transfer state
in which one electron was transferred from the cluster to the NO molecule
was used as a representative electronically excited state.
The potential energy surfaces were determined as a function of the
NO center of mass distance from the surface and the polar orientation
of the molecule with respect to the surface normal. Figure~\ref{NONiO_PES} 
shows the excited state potential.

The dynamics of the laser-induced desorption of NO from NiO(100) were
simulated by the jumping wave packet method in which the molecular
distance from the surface and the polar and azimuthal orientation
of the molecule were considered explicitly. The laser-induced electronic
excitation was modelled by putting the three-dimensional ground
state wave function of the electronic ground state onto the electronically
excited PES. Thereby the center of the wave function is located at
the Franck-Condon point FC shown on Fig.~\ref{NONiO_PES}. The minimum
of the excited state potential is closer to the surface than in the
ground state potential leading to an acceleration of the wave packet
towards the surface. This is called an Antoniewicz desorption
scenario \cite{Anton80}. The wave packet is then propagated
for a certain residence time on the excited state potential
before it is transferred in a second Franck-Condon transition
back to the electronic ground state. Molecules that have gained
enough kinetic energy after the Franck-Condon transitions are
able to desorb. The finite residence time is due 
to the interaction of the excited state with the electronic
spectrum of the substrate. The coupling strength of this interaction
that causes the quenching of the electronic excitation is still
unknown. Hence the mean residence time, the so-called {\em resonance} time
$\tau_r$, enters as an adjustable parameter in the dynamics simulation. 
A value of $\tau_r = 24$~fs has been 
chosen which yields a desorption probability of 3.3\% in agreement
with typical experimental data \cite{Klue98PRL,Mull92}.

Since the wave packet calculation are performed within a restricted
dimensionality, quantitative agreement with the experiment \cite{Mull92}
cannot be the ultimate goal of this theoretical study. More important
is to gain a qualitative understanding of experimental trends.
And indeed this study provides such a qualitative concept.
The excited state potential has a complicated shape as a function of 
the polar orientation of the molecule. Two possible pathways are sketched 
in Fig.~\ref{NONiO_PES} along which partial wave packets can propagate.
This bifurcation of the wave packet causes a bimodality in the
calculated velocity distribution of desorbing molecules. This bimodality has
also been found in the experiment \cite{Mull92}. 
Hence the {\it ab initio} treatment of the laser-induced desorption 
relates qualitative trends in the experimental results to
microscopic details of the potential energy surfaces.

\section{Conclusions and Outlook}

In this review {\it ab initio} dynamics studies of reactions
on surfaces have been presented. In the last years the interaction between
electronic structure calculations on the one side and dynamical calculations
on the other side has been very fruitful. The availability of high-dimensional
reliable potential energy surfaces has challenged the dynamics community
to improve their methods in order to perform high-dimensional dynamical 
studies on these potentials. Now quantum studies of the dissociation of 
hydrogen on surfaces are possible in which all six degrees of 
freedom of the molecule are treated dynamically. In this review I have
tried to show that this achievement represents an important step forward 
in our understanding of the interaction of molecules with surfaces.
Not only the quantitative agreement with experiment is improved, but
also important qualitative concepts emerge from these high-dimensional
calculations.

However, what are the next steps to be taken in the application
of {\it ab initio} dynamics studies?
The six-dimensional dynamical treatment of hydrogen dissociation represents
the achievement of a goal that has long been pursued. But 
due to the unfavorable scaling of the quantum dynamical methods with the
number of channels there will probably be no seven- or eight-dimensional
quantum calculations in the near future (although there are nevertheless
some promising approaches \cite{Bre96ZPC}). Neither will there soon be 
a six-dimensional quantum treatment of molecules heavier than hydrogen 
for the same reason. But for these heavier molecules quantum effects are 
not so important so that a classical treatment of the dynamics will be
sufficient.

The important class of oxidation reactions, for example, will certainly
soon be the subject of {\it ab initio} studies in which the dynamics
is treated classically. In {\it ab initio} molecular dynamics simulation 
also the energy transfer to substrate phonons can be taken into account by 
explicitly including the upper layers of the surface in the trajectory
calculations. Still the quantum treatment of dissipation is an important
goal because it would allow the energy transfer to phonons and to
electron-hole pairs to be treated on an equal footing. This issue leads
over to the treatment of electronically non-adiabatic processes.
In this important field both the determination of excited state potentials
as well as the description of non-adiabatic dynamics still represent a 
great challenge where much progress has to be made.

Finally it should be mentioned that there is still room for low-dimensional
model studies. In order to find out how the reaction dynamics depends
on specific details of the PES it is usually very instructive to perform
dynamical simulations in which certain features of the PES are changed
in a controlled fashion within a restricted dimensionality. However,
without any {\it ab initio} calculations it is often impossible to
even guess the general shape of the PES for a more complicated reaction.
This confirms the important role that {\it ab initio} dynamics calculations
play for the advancement of our understanding of reactions on surfaces,
even for model studies.

\section*{Acknowledgements}

I am very grateful to all my collegues and coworkers who have 
contributed to the results presented in this review. 
I would like to mention Michel Bockstedte, Thomas Brunner, 
Bj{\o}rk~Hammer, Peter~Kratzer, Eckhard~Pehlke, Ralf~Russ, 
Ching-Ming Wei, Steffen~Wilke, and my supervisors
Helmar Teichler, Wilhelm Brenig and Matthias Scheffler.
In particular, without the discussions and the close ongoing collaboration
with Wilhelm Brenig and Matthias Scheffler this work would have been
impossible. I would also like to thank them and Peter Kratzer and
Paolo Ruggerone for their critical reading of the manuscript.
Special thanks go to Peter Kratzer and Thorsten Kl{\"u}ner 
for sharing their results with me prior to publication.

\end{document}